\documentclass[twoside,11pt]{article}
\usepackage{jair, theapa, rawfonts}

\usepackage[T1]{fontenc}
\usepackage{graphicx}
\usepackage{comment}
\usepackage{soul}
\usepackage{xcolor}
\usepackage{booktabs}
\usepackage{multirow}
\usepackage{arydshln}
\usepackage{url}
\usepackage{tikz}
\usepackage{subcaption}
\usepackage{enumitem}
\usepackage{mathtools}
\usepackage{amsmath}
\usepackage{bm}
\usepackage{adjustbox}
\usepackage{amssymb}

\definecolor{CPI}{HTML}{19171A}
\definecolor{FdI}{HTML}{0C416E}
\definecolor{LE}{HTML}{00A01E}
\definecolor{FI}{HTML}{2491D5}
\definecolor{M5S}{HTML}{FFD600}
\definecolor{PD}{HTML}{FF9D00}
\definecolor{+E}{HTML}{EC208F}
\definecolor{PRC}{HTML}{E00916}

\definecolor{lightyellow}{cmyk}{0,0,0.50,0}
\sethlcolor{lightyellow}

\newcommand*\unsupervised[1][1ex]{\tikz\draw (0,0) circle (#1);} 
\newcommand*\semisupervised[1][1ex]{%
  \begin{tikzpicture}
  \draw[fill] (0,0)-- (90:#1) arc (90:270:#1) -- cycle ;
  \draw (0,0) circle (#1);
  \end{tikzpicture}}
\newcommand*\supervised[1][1ex]{\tikz\fill (0,0) circle (#1);}

\setlist[description]{%
  font={\normalfont\sffamily}%
}

\jairheading{73}{2022}{633-672}{06/2021}{02/2022}
\ShortHeadings{Fine-Grained Prediction of Political Leaning on Social Media}
{Fagni \& Cresci}
\firstpageno{633}

\begin{document}

\title{Fine-Grained Prediction of Political Leaning on Social Media\\with Unsupervised Deep Learning}

\author{\name Tiziano Fagni \email tiziano.fagni@iit.cnr.it \\
       \name Stefano Cresci \email stefano.cresci@iit.cnr.it \\
       \addr Institute of Informatics and Telematics (IIT)\\
       National Research Council (CNR)\\ via G. Moruzzi 1, 56124 Pisa, Italy}
\maketitle


\begin{abstract}
Predicting the political leaning of social media users is an increasingly popular task, given its usefulness for electoral forecasts, opinion dynamics models and for studying the political dimension of polarization and disinformation.

Here, we propose a novel unsupervised technique for learning fine-grained political leaning from the textual content of social media posts. Our technique leverages a deep neural network for learning latent political ideologies in a representation learning task. Then, users are projected in a low-dimensional ideology space where they are subsequently clustered. The political leaning of a user is automatically derived from the cluster to which the user is assigned. We evaluated our technique in two challenging classification tasks and we compared it to baselines and other state-of-the-art approaches. Our technique obtains the best results among all unsupervised techniques, with \textit{micro F1} $= 0.426$ in the 8-class task and \textit{micro F1} $= 0.772$ in the 3-class task. Other than being interesting on their own, our results also pave the way for the development of new and better unsupervised approaches for the detection of fine-grained political leaning.
\end{abstract}

\section{Introduction}
\label{sec:intro}
Since the advent of Facebook and Twitter, politicians have had an increasing online presence in order to reach out to as many potential electors as possible. As of today, digital campaigning (including social media) has become mandatory, as people are massively consuming political content from social platforms\footnote{\url{www.journalism.org/2018/09/10/news-use-across-social-media-platforms-2018/}}. 
Recently, 20\% of interviewed social media users admitted to have changed their minds about a political issue because of something they read on social media\footnote{\url{www.pewinternet.org/2016/10/25/the-political-environment-on-social-media/}}. 
Political activity on social media is also positively correlated to offline political activism (e.g., attending offline political events)~\shortcite{vaccari2015political}. Politically-interested users are keen to know the stance of their friends, to read about candidates and campaigns, and to discuss pressing issues and election results~\shortcite{grvcar2017stance,tucker2018social}. In spite of the relatively small readership of online platforms compared to that of traditional media (e.g., TVs, newspapers, and radio channels), the sociopolitical relevance of social media is still massive. In fact, second-order effects -- typical of complex systems -- allow for significant portions of the political social media content to be discussed also on traditional media, thus somehow still making it into the minds of people who don't even use social media at all~\shortcite{benkler2017study}. 

Given this picture, it comes with little surprise that the task of learning the \textit{political leaning of social media users} recently received a surge of attention. In literature, this task is also referred to as political \textit{stance}, \textit{ideology}, \textit{polarity} or \textit{alignment} prediction. 
Firstly, it represents a natural extension to the early efforts by social and political scientists at this task. In fact, ideology lies at the core of many theories in political science and has long been used to investigate individual behavior and preferences, governmental relations, and links between them~\shortcite{bond2015quantifying}. Traditional estimates are based on explicit preferences, such as roll-call votes, co-sponsorship records, and records of financial contributions to political campaigns. However, these data are typically available only for a few political figures (e.g., roll-call votes) or for a limited number of ordinary individuals, they are hard to acquire, and they are made available or updated infrequently. These limitations make fine-grained, continuous, large-scale analyses of political preferences challenging, if not outright infeasible. Conversely, social media represent a trove of both explicit and structured (e.g., likes and social relationships), as well as implicit and unstructured (e.g., text), data about the habits and preferences, including political ones, of millions of users. As such, many social and political scientists recently turned their attention to political analyses on social media -- e.g., by estimating political leaning from social media data and by comparing such estimates with more traditional ones~\shortcite{tucker2018social}. Meanwhile, also computer scientists found value in learning users political leaning, for a myriads of goals, such as: to forecast the outcome of elections~\shortcite{tumasjan2010predicting,ahmed2016tweets}; to estimate accurate priors for models of opinion diffusion~\shortcite{dandekar2013biased,mas2013differentiation}; to measure and mitigate online polarization~\shortcite{wong2016quantifying,garimella2017reducing,nizzoli2020coordinated}; to measure the effects of information operations, disinformation campaigns and propaganda~\shortcite{nikolov2020right,tardelli2021detecting,cinelli2020limited,ferrara2020misinformation}; to explore the political dimension of bad actors, such as social bots and trolls~\shortcite{hegelich2016social,rizoiu2018debatenight,luceri2019red,yan2020asymmetrical,cresci2020decade}.

Existing approaches to the prediction of political leaning mainly focus on analyzing only the social or interaction networks~\shortcite{garimella2016quantifying,wong2016quantifying}, or only the content of shared messages~\shortcite{pla2014political,di2018content,yan2019congressional,preoctiuc2017beyond}, with few exceptions where content and networks are simultaneously considered~\shortcite{lahoti2018joint,aldayel2019your}. Network-based approaches are grounded on the assumption that ideologically-similar users are likely to interact with, or to follow, each other. A first limitation arises when this assumption is violated -- namely, in all those cases where like-minded users never interact, or in those equally-frequent cases where opposing users interact (e.g., to argue or to convince each other). There also exist users that do not follow others, or that follow a very limited number of accounts, which inevitably complicates network-based approaches. Notable examples of this kind are \textit{@POTUS} in the US and media outlets/journalists that do not follow other accounts, for neutrality reasons, but that represent interesting subjects of political leaning analyses. Another limitation involves the large amounts of data needed for the analysis (e.g., the social or interaction graph), which are seldom promptly available. Content-based approaches are instead mainly limited by the intrinsic difficulty of processing natural language, and by the need for large corpora of manually-annotated messages and language-specific resources. Moreover, the majority of existing solutions adopt supervised approaches, which have been shown to lack generalizability and to suffer from the limited availability of comprehensive and reliable ground-truth datasets~\shortcite{cohen2013classifying}.


\subsection{Our Approach}
Our goal in this work is that of developing an unsupervised content-based technique for predicting the political leaning of social media users. We will focus on two different tasks: (i) the prediction of the preferred political party of a user (fine-grained task), which in political science literature is typically referred to as party identification; and (ii) the prediction of its political pole (coarse-grained task). For bipolar systems, the latter task simply involves the prediction of left-right ideology, which for US data is typically measured in a continuous, one-dimensional space, with techniques such as the well-known DW-NOMINATE~\shortcite{poole1985spatial}. Notably, labels obtained for the two tasks represent different user traits and should not be equated or used interchangeably. For instance, the difference between the preferred party and the ideological position of a user in the left-right scale is straightforward when considering the shifts that parties exhibit between different elections~\shortcite{busch2016estimating}. This is particularly true for the application and evaluation scenario of our work: the tripolar Italian political system~\shortcite{pasquino2019state}. Nonetheless, we are interested in evaluating the efficacy of our proposed method in solving each of the two tasks separately.

In contrast with previous work, where political \textit{ideology} and \textit{leaning} were considered as synonyms, here we make an important distinction. By drawing upon definitions from the Oxford English Dictionary, we define \textit{ideology}\footnote{\url{https://www.lexico.com/en/definition/ideology}} as a latent set of concepts that forms the basis of a user's political preferences. Instead, we define \textit{leaning}\footnote{\url{https://www.lexico.com/en/definition/leaning}} as the practical political preferences of a user (e.g., its preferred party). Our approach for predicting political leaning, independently on the task (i.e., the desired fine or coarse prediction granularity), directly stems from the previous definitions. In fact, we first adopt an unsupervised approach to learn informative political representations of social media users. We then project users into a lower-dimensional space derived from their latent representations, which corresponds to the political ideology space. Finally, we leverage the topology of the political ideology space to infer the political leaning of each user. As such, our predicted leanings strictly depend on the latent ideologies learned for every user.


\subsection{Contributions}
\begin{figure*}[t]
    \centering
    \includegraphics[width=\textwidth]{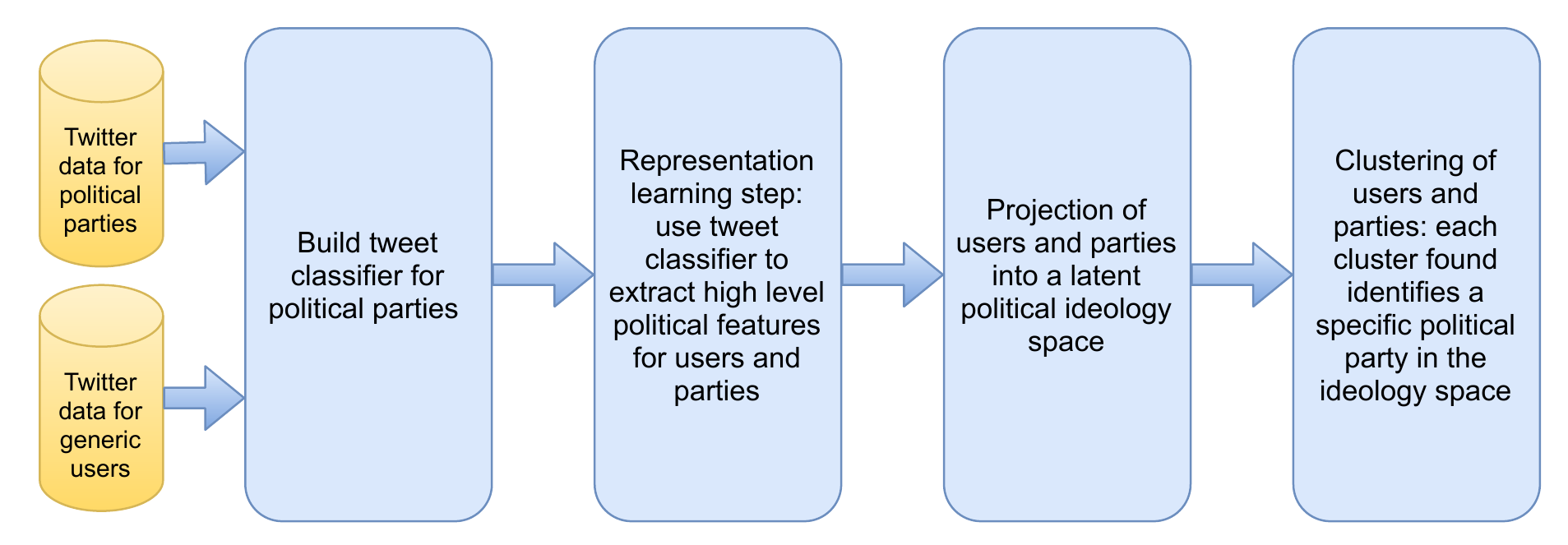}%
    \caption{Outline of our proposal.}
    \label{fig:proposal-outline}
\end{figure*}

Operationally, we propose a novel unsupervised solution for estimating the political leaning of social media users that is able to overcome the main limitations of previous approaches. Our method follows the scheme shown in Figure~\ref{fig:proposal-outline}.
Our solution initially leverages a deep neural network for learning latent users representations. Then, we feed these representations to a UMAP model in order to project and position users in a latent political ideology space. Finally, we leverage properties of the ideology space to infer the political leaning of every user, via clustering. We evaluate our proposed method and those used for comparisons on two challenging tasks. Specifically, we learn both fine-grained (i.e., party-level) and coarse-grained (i.e., pole-level) political leaning of Twitter users. Our solution achieves state-of-the-art results in both tasks, compared to existing unsupervised techniques. Specifically, it achieves $F1 = 0.43$ and $F1 = 0.77$ when predicting fine- and coarse-grained leanings, whereas other unsupervised techniques and baselines achieve $F1 \leq 0.35$ and $F1 \leq 0.71$, respectively. Our technique is exclusively based on the textual content of user-generated social media posts. However, despite exploiting solely this noisy data source, it achieves performances that are comparable or even better than techniques that leverage cleaner signals (e.g., social relationships and interactions such as retweets and likes). This makes our technique particularly valuable since it obtains state-of-the-art performance without the need for gathering explicit user preferences or data-demanding network representations. In addition, by adopting an unsupervised deep learning approach, we are also language-independent, we avoid the need for manually-annotated corpora and linguistic resources, and we improve the generalizability of our results with respect to the traditional supervised approaches that are intrinsically limited by the availability of accurate and extensive ground-truth datasets~\shortcite{cohen2013classifying,cresci2020decade}.

Our main contributions can be summarized as in the following:
\begin{itemize}
    \item We provide a state-of-the-art unsupervised method for learning both fine-grained and coarse-grained political leaning of social media users.
    \item Our nuanced solution disentangles the sub-tasks of learning latent political ideologies from that of inferring political leanings, which were mixed and overlapping in previous works.
    \item We demonstrate the usefulness of unsupervised deep learning and projection with UMAP, to accurately position users within a latent ideology space.
    \item We show the profitability of leveraging the topology of the learned ideology space to infer political leaning via clustering.
\end{itemize}

\subsection{Reproducibility}
Our data are publicly available for scientific purposes\footnote{\url{https://doi.org/10.5281/zenodo.5793346}}.

\subsection{Roadmap}
The remainder of the paper is organized as follows. In Section~\ref{sec:relwork} we discuss previous works on the prediction of political leaning from social media. Then, before presenting our solution, in Section~\ref{sec:prelim} we outline the political context in which our study is positioned and we provide details about our dataset. Section~\ref{sec:ideology} describes our deep learning approach for learning latent political ideologies of social media users. In Section~\ref{sec:prediction} we discuss our approach for positioning users in a latent political ideology space, and for inferring their political leaning. Experiments\footnote{Throughout the manuscript we use the term ``experiment'' with its conventional meaning in computer science -- that is, an analysis, measurement, or evaluation campaign. This is different from its meaning in other disciplines (i.e., the social sciences) where experimental approaches involve treatments or interventions and are opposed to observational ones.} and results are presented in Section~\ref{sec:results}, while Section~\ref{sec:conc} draws conclusions and highlights promising directions for future work.

\section{Related Work}
\label{sec:relwork}
In this section we briefly survey extant literature for the prediction of political leaning. We split previous works based on the information used to make predictions.

\subsection{Content-Based Approaches}
Among the first approaches at this task are those solely based on the analysis of the textual content of messages. \shortciteA{pla2014political} investigated the use of sentiment analysis features. They trained a supervised classification model capable of labeling users based on their coarse-grained leaning -- namely, as either left-leaning, right-leaning, center-leaning or undefined. Similarly,~\shortciteA{di2018content} leveraged a set of linguistic syntactic features in a supervised classification task. The goal of their system was that of learning the political preference of Twitter accounts towards the 4 main parties in Italy. Despite focusing on fine-grained (i.e., party-level) predictions,~\shortciteauthor{di2018content} worked with only 4 parties, instead of the 8 considered in our present work.

The previous works are representatives of a rather large body of work based on supervised content classification. Results obtained by these systems are however disputed by~\shortciteA{cohen2013classifying}, since they tend to overestimate performances by focusing on politically active users (instead of \textit{normal} or politically \textit{inactive} users) and since their classification performances rapidly plummet when applied outside of the narrow range of examples used for training the systems. Similar results were also recently obtained by~\shortciteA{yan2019congressional}, who evaluated the generalizability of text-based supervised systems for classifying partisanship and political ideology. Specifically, the authors built 3 datasets derived from the US Congressional Record, polarized media websites, and political wikis. Then, they trained a set of supervised classifiers on a dataset and they evaluated their performance in classifying texts from the other datasets. Among the supervised algorithms used for text classification are logistic regression as well as deep learning-based classifiers such as Marginalized Stacked Denoising Autoencoders and Semi-Supervised Recursive Autoencoders. Results show the difficulty of supervised and semi-supervised systems in generalizing from one dataset to another, thus motivating research and experimentation with unsupervised approaches.

Another major drawback of supervised classification is that political leaning is typically provided as a discrete (e.g., binary) variable. A first improvement over these works was done by~\shortciteA{preoctiuc2017beyond}, who predicted the political orientation of Twitter users on a 7 point scale ranging from ``very conservative'' to ``very liberal'', with several points reserved for moderate users. They leveraged several features extracted from the tweets posted by the analyzed users, including features derived from LIWC, sentiment, topics, named entities and word2vec, and the prediction was performed with simple supervised classification algorithms (e.g., logistic regression). Conversely, more recent works moved towards unsupervised approaches. \shortciteA{kulshrestha2017quantifying} proposed a system where the leaning is obtained by measuring the similarity between the topic vectors of users, with those of known seed democrats and seed republicans. The political leaning was provided for each user in the $[0, 1]$ continuous range. This system is unsupervised, however it requires known sets of seed users, raising the question as to how to obtain such sets. Moreover, an additional challenge to face when developing systems for predicting continuous (rather than binary or crisp) leanings, is the lack of ground-truth values for training or evaluating the system.

\subsection{Network-Based Approaches}
Approaches purely focused on network characteristics currently represent only a minority of existing works. \shortciteA{barbera2015birds} built a Bayesian spatial model of the Twitter social network that is based on homophilic network properties. The political leaning of each user is determined via Ideal Point Estimation. Similarly,~\shortciteA{bond2015quantifying} exploited user likes to Facebook pages to obtain estimates of political ideology for both parties, politicians, and ordinary users. Estimates are computed via Singular Value Decomposition (SVD) of an agreement matrix, which corresponds to a normalized adjacency matrix derived by projecting the bipartite matrix of user likes to parties onto the set of parties. Instead,~\shortciteA{wong2016quantifying} computed political leaning by solving a convex optimization problem. By leveraging Twitter data, the objective function embeds signals derived from both the analysis of retweeting behaviors and features of the retweet networks. These previous works are unsupervised and provide leaning estimates in the $[0, 1]$ continuous range. Notably, these works, as well as all others that output one-dimensional scores, can only be applied to bipolar systems (e.g., to binary prediction tasks). This means that they are not suitable for application to the detection of fine-grained political leaning, a task that demands the prediction of multiple classes (i.e., the possible political parties), nor to the detection of coarse-grained political leaning in those systems that have more than two poles. An example of the latter is the current Italian political system~\shortcite{pasquino2019state}, to which we apply our proposed methodology. The usefulness of the so-called left-right scale, operationalized as the $[-1, 1]$ or $[0, 1]$ continuous range, is also questioned by~\shortciteA{bauer2017left}, who found that different individuals assign different meanings to the ``left'' and ``right'' concepts. As such, estimates based on a unique left-right scale for all individuals risk being biased and inaccurate. More broadly,~\shortciteauthor{bauer2017left} also raised the issue of self-reports, such as those obtained from survey respondents, as a ground-truth for training automated systems. In fact, many recent studies uncovered severe biases in self-reports, which motivates research on alternative means of obtaining ground-truth measurements~\shortcite{bastick2021would,verbeij2021accuracy}.

A recent interactions-based state-of-the-art unsupervised approach is presented by~\shortciteA{darwish2020unsupervised}. Authors built user representations based on the users they retweeted. Then, they experimented with several projection and dimensionality reduction techniques, such as t-SNE and UMAP. Finally, they clustered projected users and labelled clusters via manual inspection. As a result of this process, each user is assigned to the label of the cluster to which it belongs. The system presented by~\shortciteA{darwish2020unsupervised} has been employed also for predicting the political bias of media outlets and famous public characters~\shortcite{stefanov2019predicting}, and to estimate the polarization of Twitter users with respect to certain debated topics and political issues~\shortcite{darwish2018kavanaugh}.

The aforementioned work is the most similar existing solution with respect to our present contribution. However, contrarily to~\shortcite{darwish2020unsupervised}, we do not explicitly exploit retweets between users, but we rather leverage the noisy textual content of their tweets. Consequently, a crucial component in our solution is the deep learning network used to learn latent user representations from tweets. In addition, we make different choices with respect to the techniques used for dimensionality reduction, projection and clustering. Finally, we automatically label clusters based on the labels of the pivots contained in each cluster, rather then with manual intervention. In our work, we also evaluate systems on a more challenging task than that tackled by~\shortciteA{darwish2020unsupervised} (e.g., binary classification), demonstrating and discussing the advantages of our solution.

\subsection{Mixed Approaches}
Another large body of work is based on a combination of content and network analysis. The advantage of simultaneously exploiting both textual content and network representations, such as those resulting from user interactions, was recently motivated and quantified by~\shortciteA{aldayel2019your}. Specifically, they found that several different dimensions of online profiles and activities can provide useful signals to predict stance and leaning. Among them, some of the most informative signals can be extracted from user posts, user interactions with other users, websites visited, and user likes to other content on the platform.

Among the first works to jointly exploit content and interaction networks is~\shortcite{conover2011predicting}. Authors exploited features derived from hashtags and from the retweet network, in a supervised binary classification task. Similarly, also~\shortciteA{pennacchiotti2011democrats} focused on supervised binary political classification. Their system is fed with features encompassing profile, tweeting behavior, linguistic, social, and interaction network information. Being based on supervised classification, both previous works still suffer from the limitations outlined by~\shortciteA{cohen2013classifying} and can only provide a dichotomic estimate of polarity. The work by~\shortciteA{lahoti2018joint} instead provided interesting advances on this task. It is a state-of-the-art unsupervised framework based on non-negative matrix factorization, which learns a shared latent space between users and content. Similarly to other already-surveyed works, political leaning is considered as a one-dimensional continuous variable in the $[0, 1]$ range. 
However, the framework can be used to model more than one variable at a time (e.g., ideology, popularity), which represents an interesting improvement over previous works. 

One of the limitations of network-based and mixed approaches is the need for explicit social relationships or user preferences (e.g., likes, retweets). In fact, it has been demonstrated that extracting these information is a data- and time-demanding task and that such information is not always available (e.g., due to platform data-access restrictions)~\shortcite{cresci2015fakefollowers}. In turn, this decreases the applicability of such techniques and hinders large-scale social media analyses. Contrarily, our proposed technique achieves comparable or better performances while only exploiting the textual content of user posts, which are readily available.


\section{Preliminaries and Data}
\label{sec:prelim}
This section provides preliminary information on the political landscape in which our analyses take place. Furthermore, it provides details on our dataset and its labeling.

\subsection{The Italian Political Landscape}
\label{sec:prelim-politics}
We focus our study on politically-active Italian Twitter users. Thus, our aim for this work is predicting the leaning of Italian Twitter users, within the current Italian political spectrum. Before delving into the details of our methodology, we first outline the Italian political landscape as of November 2020.

The last Italian general elections were held in March 2018, and resulted in the populist party Five-star Movement (\textsf{M5S}) winning the election with 32.7\% votes, followed by the center-left Democratic Party (\textsf{PD}) with 18.7\% votes and the far-right League (\textsf{LE}) party that obtained 17.4\% votes. Despite receiving slightly more votes than \textsf{LE}, \textsf{PD} is considered one of the losers of the election, since it dropped from 40.8\% votes received at the 2014 European elections, to 18.7\% in 2018. The last major Italian party is the center-right Forward Italy (\textsf{FI}) that obtained 14.0\% votes. Based on this outcome, a coalition government was formed in May 2018 by \textsf{M5S} and \textsf{LE}. This lasted until August 2019, when a government crisis initiated by \textsf{LE} led to the formation of a new coalition government in early September. This government, which is still in charge at the time of writing, is led by \textsf{M5S} and \textsf{PD}, together with other minor parties\footnote{\url{https://en.wikipedia.org/wiki/Conte_II_Cabinet}}. The peculiarity of the current Italian political landscape is represented by the populist and anti-establishment \textsf{M5S}, whose members refuse to position in the traditional left-right bipolar paradigm since they regard \textsf{M5S} as a \textit{non-party}\footnote{\url{https://en.wikipedia.org/wiki/Five_Star_Movement}}. As a consequence, the coarse-grained Italian political landscape is a tripolar system consisting of right-leaning parties, left-leaning parties, and the \textsf{M5S}~\shortcite{pasquino2019state}. Notably, carrying out predictions of political leaning in a tripolar system has implications on the techniques used for the analysis, since some of the existing ones have been specifically designed for bipolar systems (e.g., left \textit{vs} right, liberals \textit{vs} conservatives, in favor \textit{vs} against a given topic).


In addition to the aforementioned parties, in this study we also consider 4 minor parties that together accounted for 8\% votes in the 2018 general elections, thus covering the whole extent of the Italian political spectrum and including both major and minor parties. Table~\ref{tab:politics} summarizes the main information, name and color conventions for all considered parties and their leaders. Henceforth, we refer to the party Twitter accounts as our \textit{pivots}, since they play an important role in the estimation of political leaning. Notably, the only preliminary data needed by our framework are (i) the pivots, and (ii) their coarse-grained leaning. 

\begin{table*}[t]
	\small
	\centering
	\begin{adjustbox}{max width=\textwidth}
	\begin{tabular}{cllllcr}
		\textbf{leaning} & \multicolumn{1}{c}{\textbf{party name}} & \multicolumn{1}{c}{\textbf{party handle}} & \multicolumn{1}{c}{\textbf{leader handle}} & \multicolumn{1}{c}{\textbf{label}} & \textbf{color} & \multicolumn{1}{c}{\textbf{\#users}} \\
		\toprule
            \multirow{4}{*}{\textsf{RIGHT}} & CasaPound Italy         & \textit{@casapounditalia} & \textit{@distefanoTW}     & \textsf{CPI}  & \tikz\draw[black, fill=CPI] (0,0) circle (.85ex); & 2,997 \\
            & Brothers of Italy       & \textit{@FratellidItaIia} & \textit{@GiorgiaMeloni}   & \textsf{FdI}  & \tikz\draw[black, fill=FdI] (0,0) circle (.85ex); & 2,507 \\
            & League                  & \textit{@legasalvini}     & \textit{@matteosalvinimi} & \textsf{LE}   & \tikz\draw[black, fill=LE] (0,0) circle (.85ex);  & 2,705 \\
            & Forward Italy           & \textit{@forza\_italia}   & \textit{@berlusconi}      & \textsf{FI}   & \tikz\draw[black, fill=FI] (0,0) circle (.85ex);  & 746 \\
        \midrule
            \multirow{1}{*}{\textsf{M5S}} & Five-star Movement      & \textit{@Mov5Stelle}      & \textit{@luigidimaio}     & \textsf{M5S}  & \tikz\draw[black, fill=M5S] (0,0) circle (.85ex); & 3,206 \\
        \midrule
            \multirow{4}{*}{\textsf{LEFT}} & Democratic Party        & \textit{@pdnetwork}       & \textit{@nzingaretti}     & \textsf{PD}   & \tikz\draw[black, fill=PD] (0,0) circle (.85ex);  & 2,377 \\
            & +Europe                 & \textit{@piu\_europa}     & \textit{@bendellavedova}  & \textsf{+E}   & \tikz\draw[black, fill=+E] (0,0) circle (.85ex);  & 4,335 \\
            & Communist Ref.          & \textit{@direzioneprc}    & \textit{@maurizioacerbo}  & \textsf{PRC}  & \tikz\draw[black, fill=PRC] (0,0) circle (.85ex); & 1,326 \\
		\bottomrule
	\end{tabular}
	\end{adjustbox}
	\caption{Information about the 8 Italian parties, and their leaders, considered in this study. Rows are grouped according to the coarse-grained political leaning, representing the tripolar Italian political system.\label{tab:politics}}
\end{table*}

\subsection{Twitter Dataset}
\label{sec:prelim-dataset}
Our aim for this work is to develop a framework for estimating political leaning in an unsupervised fashion (i.e., with no manual labeling involved). To combine the strengths of labeled datasets (e.g., rich, high-quality data) with those of unsupervised approaches (e.g., generalizability, no bias or errors due to manual labeling), our desiderata is to acquire a dataset that is \textit{implicitly} labeled, with respect to political alignment. We met our desiderata by leveraging favorited (i.e., liked) tweets, and by considering political likes as proxies for political leaning. Other options, also adopted in some previous works, could have involved the exploitation of retweets or follower relationships to political parties. However, we consider likes to be stronger indicators of political preference~\shortcite{aldayel2019your}.

Operationally, we first crawled the Twitter timelines of our pivots. Then, for each collected tweet, we obtained a list of users that liked that tweet. At the end of this process we obtained a bipartite graph linking 20,199 users to our 8 considered parties, based on explicit user likes to party tweets. The number of users that liked at least one party tweet is reported in the last column of Table~\ref{tab:politics}, for every party. We completed our data collection by crawling the most recent 200 tweets from the timelines of all 20,199 users, which resulted in more than 3.6M tweets, in total. When building the dataset, we only included users whose timeline contained at least 25 tweets. For each user, we collected at most up to 200 tweets. This data collection process roughly covered the months of August to early October 2019. On average, user timelines include 179.3 tweets, evenly distributed during our data collection period. Finally, we performed a stratified sampling to split our dataset into a training (90\% -- 18,169 users), a validation (3\% -- 604 users) and a test (7\% -- 1,426 users) partition. As a result of our splitting strategy, the distributions of parties and poles across the 3 data partitions are comparable.


\begin{figure*}[t]
    \centering
    \includegraphics[width=0.7\textwidth]{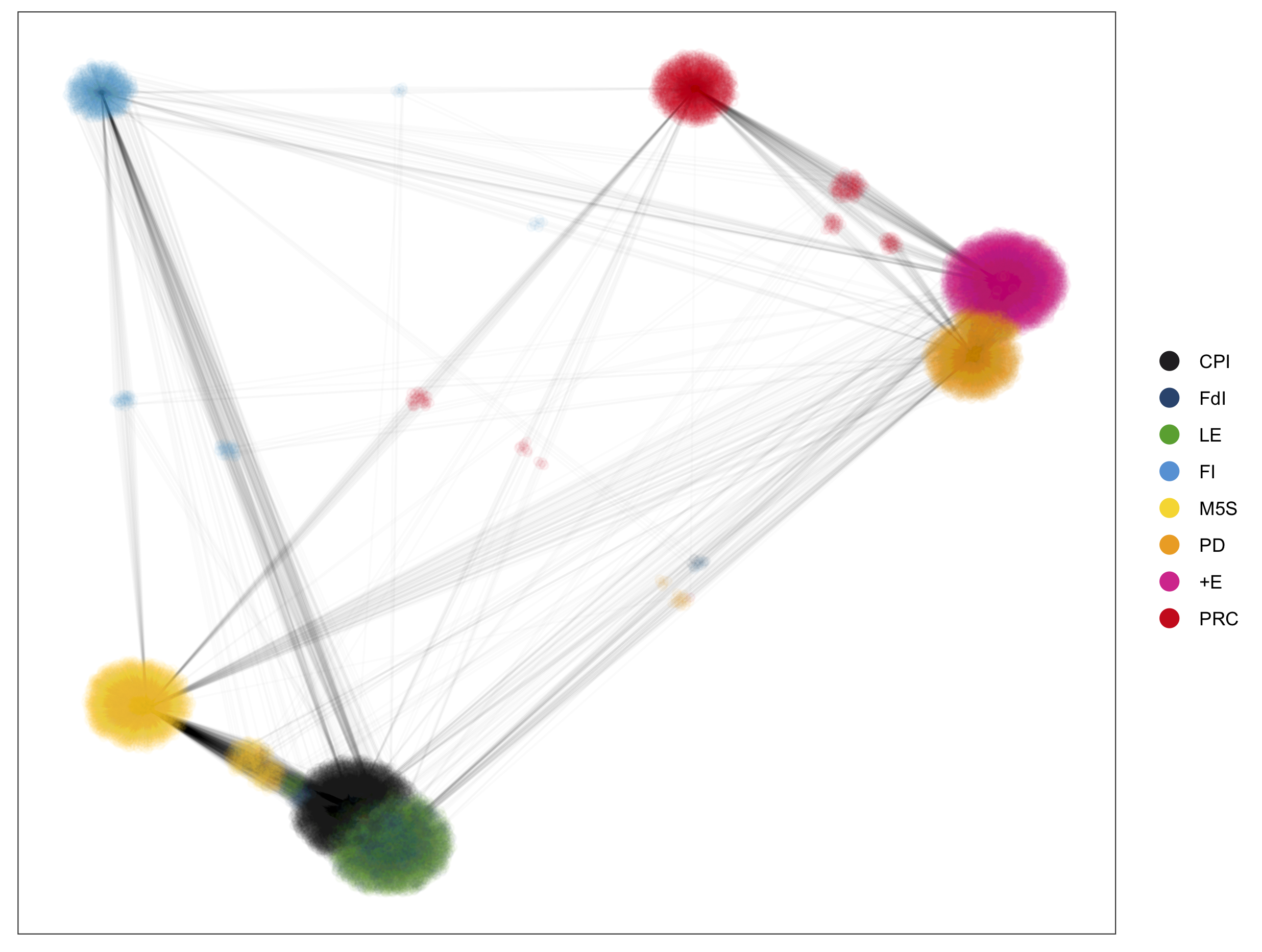}
    \caption{Louvain clustering of the weighted user-similarity network. Edge weights are based on user likes to party tweets. Clusters are color-coded and each cluster is associated to a political party. User labels resulting from this clustering are used as ground-truth for evaluating predictions of user political leaning.\label{fig:likes-clustering}}
\end{figure*}

\subsection{Ground-Truth Labeling}
\label{sec:prelim-ground-truth}
Since we do not know the preferred party of the users in our dataset, we obtain a ground truth for our task by leveraging user likes to party tweets. Specifically, we first build the bipartite graph of users and party tweets, where links between nodes represent user likes to party tweets. Next, we project the bipartite graph onto the subset of user nodes, obtaining a weighted, undirected user-similarity network. Links in this network represent similarity between users. In order to build this network and to compute the similarity between users, we adopt a simple weighting scheme based on the frequency of common associations in the bipartite graph. In other words, the similarity between two users is measured as the number of tweets liked by both users. Finally, we cluster users in this network with the Louvain community detection algorithm~\shortcite{blondel2008fast}. Each user is then labeled with the political party corresponding to the cluster it belongs to. Figure~\ref{fig:likes-clustering} shows the clustered user-similarity network derived from our dataset. Clusters are color-coded and determine the ground-truth label for each user. As shown, in this representation user clusters are sharply defined. The vast majority of users only has edges connecting to other users of the same cluster, with only a few edges connecting users across different clusters. In turn, this implies that user likes to party tweets are a very strong signal of political alignment. In the following, we describe our approach for the challenging task of inferring user political leaning from tweets, which represent a much more noisy signal than likes.


\section{Learning Latent Political Ideologies}
\label{sec:ideology}
Determining users political leaning from the analysis of the content posted on social media is a challenging task. One challenge stems from the need to find a clever way to focus the analysis on politically-relevant content only. Indeed, a typical user's timeline is filled with posts related to several different topics (e.g. sport, spare time, work, politics) that embrace all aspects of the user's life. The first difficult step is therefore related to splitting the relevant contents (i.e., those related to politics) from the rest of the messages that, within this context, simply represent noisy and unhelpful data for inferring users political leaning. A second critical aspect, given a post covering political topics, is to properly measure how such message is politically close to the typical ideology of a specific party or pole. This measure can predict how much the post is in agreement toward a specific ideology, so having access to enough messages in a user's timeline that convey this information can contribute to accurately estimate its final political leaning.

\begin{figure*}[t]
    \centering
     \includegraphics[width=\textwidth]{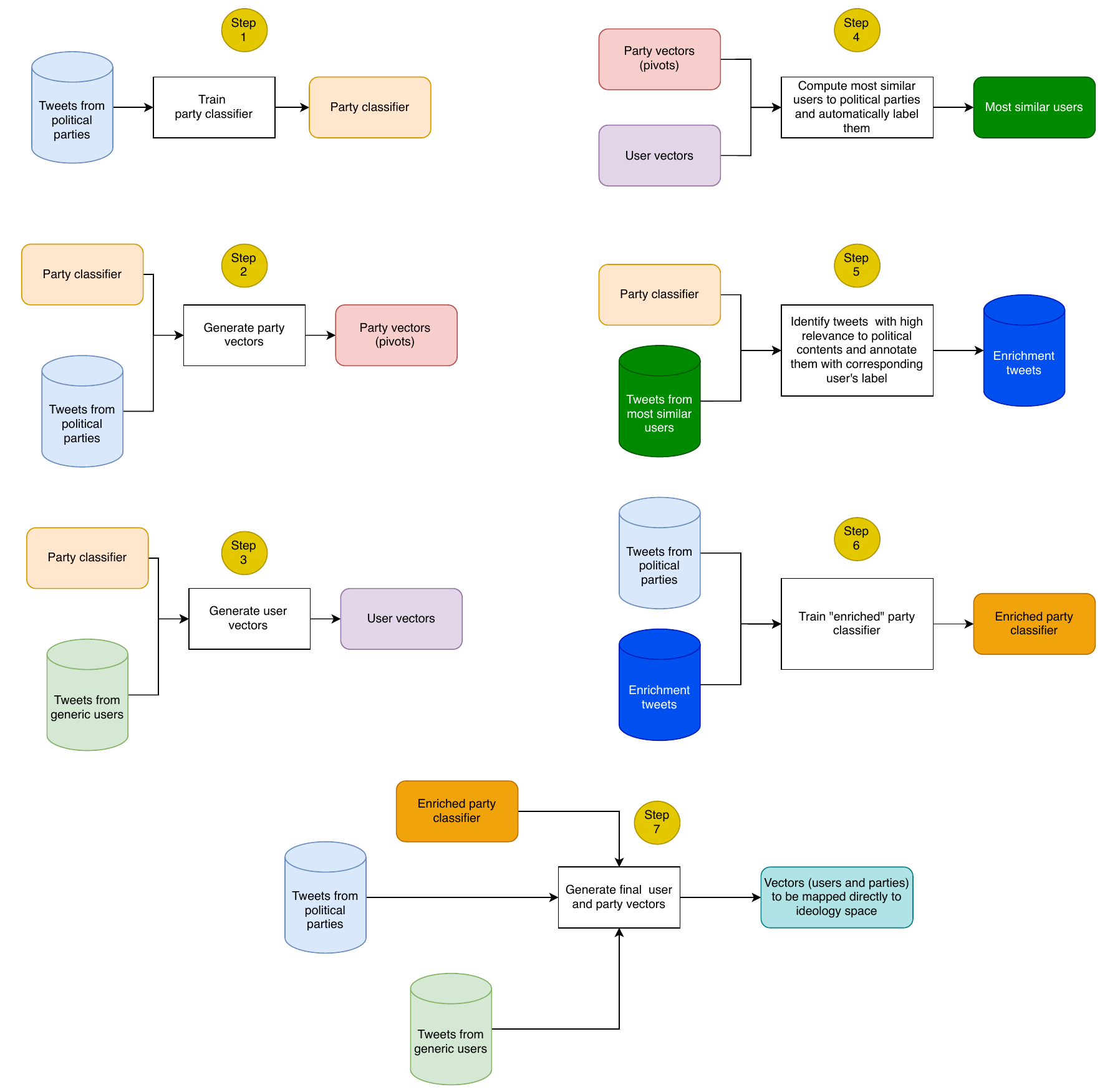}%
    \caption{High-level overview of the proposed unsupervised strategy to map users to a latent political ideology space.}
    \label{fig:learning-political-ideology-overview}
\end{figure*}

We sorted out both these critical issues by proposing a novel unsupervised process organized in seven high-level steps, shown in Figure~\ref{fig:learning-political-ideology-overview}. In step 1 we build an automatic tweet classifier for assessing if a tweet has been produced by a certain political party. Details on how the classifier is trained are given in Section~\ref{sec:party-classifier}. In steps 2, 3, and 4 we leverage the classifier to compute vector representations for users and parties. We exploit representations of users and parties to identify a subset of users that are particularly similar to the considered parties. These steps of our methodology are described in Section~\ref{sec:user-vector}. In step 5 we automatically analyze the tweets from the subset of users whose representations are similar to those of the parties. In particular, for each of such users we select a subset of his/her tweets that conveys explicit political opinions. In step 6 we use these user-generated tweets as additional training examples in a second training phase of our tweet classifier, since they represent political tweets with different characteristics than those already seen by the classifier (i.e., those obtained from the official party accounts rather than from ordinary users). The final classifier is used in step 7 to compute the final vectors of users and parties. Each computed vector corresponds to the latent representation of a user. In other words, learned vectors allow to position users in a shared latent political ideology space. Steps 5, 6, and 7 are described in Section~\ref{sec:enrichment}. Finally, in Section~\ref{sec:prediction} we describe how we leverage the relative positions of users in the latent political ideology space to infer the preferred party for each user.

\subsection{Predicting the Political Relevance of a Tweet}
\label{sec:party-classifier}
In this step we are interested in measuring the degree of agreement of a relevant political tweet with the typical political ideology of each party involved in this study. Before describing how this step works, it is helpful to define what a relevant tweet is. We deem a tweet to be \textit{politically relevant} if it expresses a subjective opinion in favour or against a specific party, a party leader, a specific person, or a political position ideologically known to be near to a certain party. In this context, possible examples of relevant tweets are unquoted retweets of tweets posted by political parties or leaders, unquoted retweets of messages of other users where they express a political opinion on something, tweets replaying to political leaders where a user shows its appreciation or a negative attitude toward the author of message.

Given a tweet, it is possible to quickly determine if the text is politically relevant by leveraging an automatic multiclass classifier which has learned, from examples of political tweets, to predict if a tweet has been produced by a specific political party. Indeed, such classifier should be able to assign not only proper labels (i.e., the most probable party that could have produced that tweet) but also to estimate a confidence in its decision which can be seen as a ``relevance level'' of the tweet with respect to political topics\footnote{The higher is the confidence score of the classifier, the higher is the probability that the tweet content is expressing something which is politically relevant.}. Such type of solution normally requires a manual annotated dataset. Here, we obtained the same result in an unsupervised way by exploiting the implicit relationship between tweets and the Twitter accounts that have produced them. In particular, we focus on the official accounts of the 8 considered parties and we use their timelines to automatically build a labeled dataset. In this way, each tweet posted by party account $P$ is labeled as generated by political party $P$ and the problem we solve is the prediction of the party that produced a tweet only based on the textual content of the tweet itself. Our approach thus resembles labeling schemes by \textit{distant supervision}~\shortcite{marchetti2012learning}. Building our dataset by focusing on party accounts also has two important practical implications that simplify solving our task. The first implication is that we are certain that the labels assigned to tweets are correct. This allows us to train a classifier on a real gold-standard thus avoiding sub-optimal solutions caused by biases and labeling errors introduced during error-prone manual labeling operations~\shortcite{misra2016seeing,pandey2019modeling}. The second and most important implication is that each considered timeline is ``clean'' (i.e., not noisy) and contains only politically-relevant tweets\footnote{It is extremely rare that an official account of a political party posts something which is not related to politics.}. Such politically-relevant tweets are typically in favour of the party, of its leader, or of some action proposed, and only seldom against another political competitor.

\begin{figure*}[t]
    \centering
    \includegraphics[width=\textwidth]{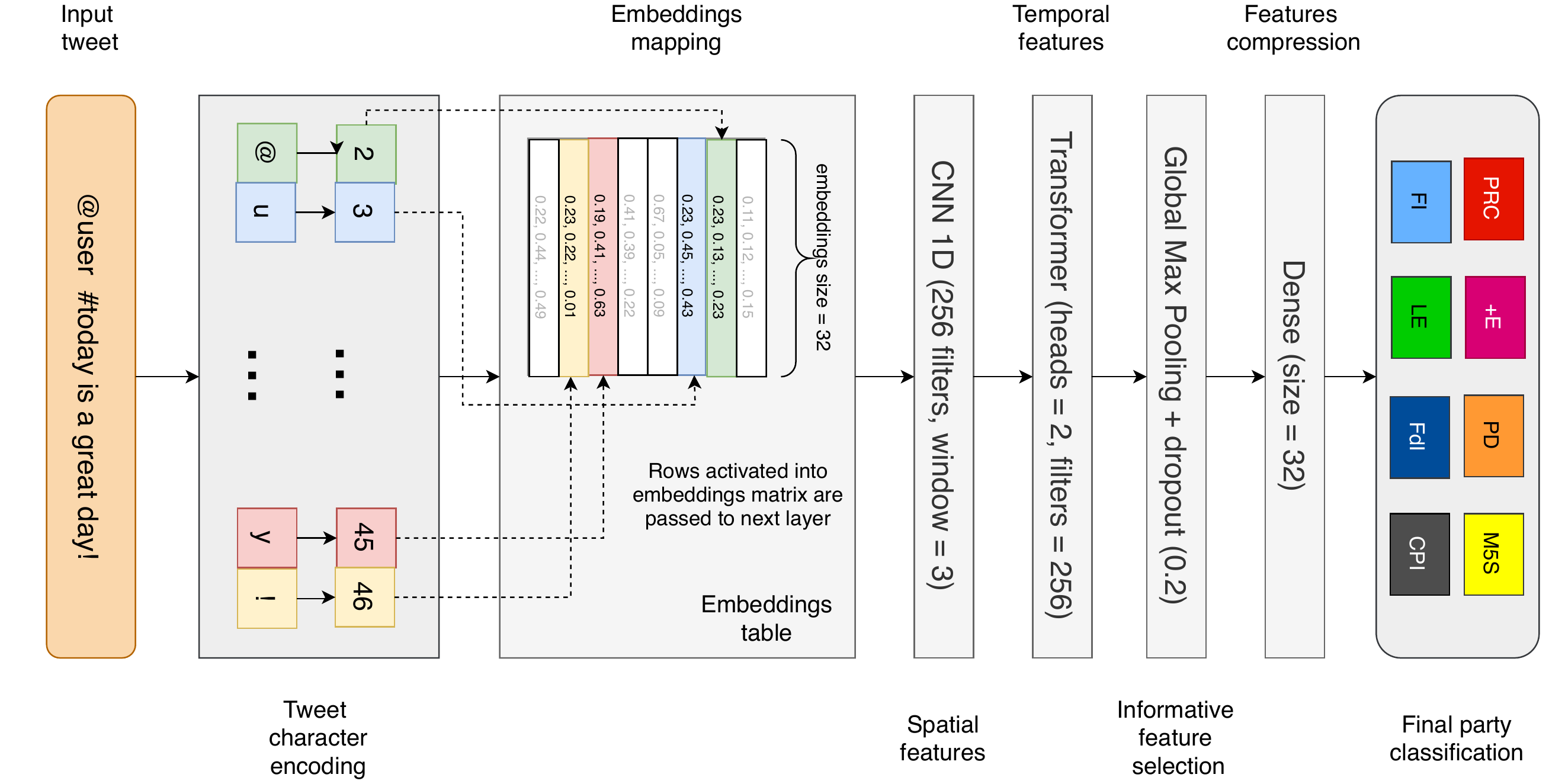}%
    \caption{Neural network architecture of our party classifier.}
    \label{fig:party-classifier-architecture}
\end{figure*}

By following the strategy described above, we built a dataset composed by all the tweets (in the form of original contents or retweets) posted in the timeline of the considered 8 political parties. For each party we selected the most recent 3,000 tweets. At the end of this process we obtained an almost balanced dataset composed of 23,791 labeled tweets. By leveraging these data, we built the political party classifier using the neural network architecture shown in Figure \ref{fig:party-classifier-architecture}. We used a character-based encoding to obtain an initial vector for each tweet. Our method uses an embeddings character table which is learned during the training phase. Each tweet vector is thus mapped into a new vector using the embeddings table and next passed to a CNN layer~\shortcite{lecun1998gradient} with the aim of extracting spatial features -- i.e., those that are invariant to the locations where they occur. This set of features is then processed by a transformer layer~\shortcite{vaswani2017attention} that extracts the most informative temporal recurrent patterns from the data. The gathered information are thus filtered and compressed before being used to produce the final classification of the tweet. 

\subsection{Extracting a Politically Relevant User Vector}
\label{sec:user-vector}
The party classifier built in the previous section can be employed to extract high level features that express the political attitude of a user with respect to all parties. In particular, by processing the entire timeline of a user with such classifier we can identify which tweets are politically relevant (i.e., tweets classified with medium/high confidence scores) and which political parties the user's opinion aligns with. More formally, let us define $T_{U} = \left\{t_{i}^{U} \; | \; i=1,...,min\left(200, |T_{U}|\right)\right\}$ as the timeline of user $U$ where $t_{i}$ is its $i$-th most recent tweet and $|T_{U}|$ is the number of tweets available in the whole timeline of $U$. Let us also declare $P = \{\textsf{PRC},\textsf{+E},\textsf{PD},\textsf{M5S},\textsf{FI},\textsf{LE},\textsf{FdI},\textsf{CPI}\}$ as the set of the 8 considered parties such that $P_{1} = \textsf{PRC}$, $P_{2} = \textsf{+E}$, and so on. We define the party classifier as the function $C$ mapping a tweet $t$ to a score vector as in the following:
\begin{equation}
C:t \in \mathbb{R}^{280} \rightarrow  \left[s_{P_1}, s_{P_2}, \ldots, s_{P_8}\right] \in \mathbb{R}^{8}
\end{equation}
where $s_{P_i}\in[0,1]$ is the score assigned by classifier $C$ to the tweet $t$ for party $P_i$. Given the timeline of user $U$, we can compute $S_{U,i,k}$ as the set of the best $k$ scores obtained for a specific party $P_i$, as in the following:
\begin{equation}
S_{U,i,k} = \left\{max_k\left\{C\left(t\right)_i\right\} \; | \; t \in T_U\right\}\in \mathbb{R}^{k}
\end{equation}
Given the previous definitions, we can finally define how to extract a politically relevant user vector:
\begin{equation}
V_{U}^{k} = \left[S_{U,1,k}, S_{U,2,k},\ldots,S_{U,8,k}\right] \in \mathbb{R}^{8k}
\label{eq:user-vector}
\end{equation}
The vector $V_{U}^{k}$ is a concatenation of the best tweet scores measured on the relevance to each party, which is indicative of the interests and the leaning shown by the user toward a specific political ideology. In this work, we fixed $k = 5$ in Equation~\eqref{eq:user-vector} based on early experimentation demonstrating this value to yield reliable measures of the degree of interest shown by a user for a specific political party. Indeed on the one hand, a larger $k$ would require the user to post a lot of political content in order not to penalize excessively the weight of a specific political stance. On the other hand, a smaller $k$ would require to have an extremely accurate party classifier.

\subsection{Unsupervised Data Enrichment to Improve Tweet Party Classification}
\label{sec:enrichment}
As for all supervised classifiers, the party classifier built in Section~\ref{sec:party-classifier} works well when analyzing tweets whose writing style is similar to that of tweets used in the training dataset. In particular, using party accounts as positive seeds in the dataset construction phase, poses some limitations to the learned classifier for correctly handling the true tweet distribution. 
Indeed, official party tweets are typically written in a clean, formal and institutional language. In addition and as previously anticipated, they also typically provide facts in support of the work of the party or of its political leader. On the contrary, political tweets from average users have different linguistic characteristics. Their writing style is informal and tweets contain abbreviations, slang, and jargon expressions. Regarding the opinions conveyed in a typical user tweet, sometimes users support a political party or leader. However, oftentimes users also express strong disagreement toward an opposing political opinion or politician. In particular, a considerable set of users tend to provide more destructive opinions (e.g., harsh comments against someone or something) than constructive ones~\shortcite{nizzoli2020coordinated}. In a few edge cases, the political opinions expressed in a user's timeline are exclusively against something or someone. Because of this, it is important to transfer such nuances to the party tweet classifier during its training phase, in order to be able to infer accurate user political ideologies.

Given these motivations, here we propose an unsupervised strategy to enrich original training data with labeled tweet examples coming from all types of users. This enrichment process is aimed at providing also negative tweet examples to the tweet party classifier, in addition to the positive tweets from the official parties, and can be summarized in the following steps:
\begin{enumerate}
    \item Obtain the vector representations for the pivot (i.e., party) accounts. This can be achieved with Equation~\eqref{eq:user-vector}.
    \item Select those training-set users that are most similar to each party account.
    \begin{enumerate}
        \item For each training-set user, we obtain its vector representation with Equation~\eqref{eq:user-vector} and we compute its cosine similarity with respect to the vector of each party.
        \item For each party, we sort users based on their similarity and we select users laying above the $99$-th percentile of the similarity distribution (i.e., the most similar ones). We automatically assign the label of the party to each user matching this condition.
    \end{enumerate}
    \item Select tweets to be used as an enrichment for training the tweet party classifier. To reach this goal, we analyze the timelines of the users selected at the previous step. For each selected user, we use the party classifier $C$ to predict the political relevance of all the tweets in the user's timeline. Then, we retain only those tweets for which $C$ yielded a score $s_{P_i} \geq Th$ for at least one party $P_i$. For large values of the threshold $Th$, this results in selecting only those tweets for which our classifier provided strong predictions. Such tweets are used as enrichment tweets in a second training phase of the classifier. The ground-truth label assigned to those tweets is that of its author, assigned at step 2.1 of this procedure. Notably, this label that we inferred in an unsupervised fashion is likely to be correct since we are considering users that are very similar to a given party in the political ideology space. Overall, this process allows to expand the training set by ingesting tweets from average users in addition to those of official party accounts, while still retaining a high confidence of the new tweets' labels.
    \item Build a new party classifier $C'$ using both the original training dataset and the enrichment data, using the same architecture shown in Figure \ref{fig:party-classifier-architecture}. As a consequence of the enriched training-set, the new classifier $C'$ is more accurate than $C$, especially with regards to informal and negative political tweets.
\end{enumerate}

\begin{table*}[t]
    \footnotesize
    \centering
    \begin{adjustbox}{max width=\textwidth}
	\begin{tabular}{lp{6cm}p{6cm}cc}
        \toprule
        & \textbf{original tweet} & \textbf{translated tweet} & \textbf{$\bm{L_c}$} & \textbf{$\bm{L_o}$}\\
        \hline
        $\surd$ & @Mov5Stelle Invece di tagliare la rappresentanza, bastava dimezzare gli stipendi. Una legge ordinaria, sicuramente più veloce come iter di una legge costituzionale. & @Mov5Stelle Instead of cutting the representation, it was enough to halve the salaries. An ordinary law, a procedure certainly faster than a constitutional law. & \textsf{PRC} & \textsf{M5S} \\
        \hline
        $\surd$ & Salvini chiede I pieni poteri (!!!), sappiatelo... & Salvini asks for full powers (!!!), you should be aware of this... & \textsf{+E} & \textsf{LE} \\
        \hline
        $\surd$ & RT @gennaromigliore: A Genova la polizia rompe le ossa al cronista \#StefanoOrigone e protegge le canaglie di \#CasaPound: bisogna sanzionare... & RT @gennaromigliore: In Genoa the police breaks the bones of the reporter \#StefanoOrigone and protects the villains of \#CasaPound: you have to punish... & \textsf{PD} & \textsf{CPI} \\
        \hline
        $\surd$ & \#CasaPound  Si arriverà davvero allo sgombero dei "fascisti del Terzo Millennio" dal palazzo di 6 piani che hanno occupato abusivamente da 15 anni nel centro di Roma? & \#CasaPound Will there really be the evacuation of the "fascists of the Third Millennium" from the 6-storey building they have squatted in the center of Rome for 15 years? & \textsf{M5S} & \textsf{CPI} \\
        \hline
        $\surd$ & @dariofrance @nzingaretti Sarebbe un governo peggio di questo. Peggio del \#pd non ci sta niente in circolazione! & @dariofrance @nzingaretti It would be a worse government than this. There is nothing around that is worse than \#pd! & \textsf{FI} & \textsf{PD} \\
        \hline
        $\surd$ & Un Altro PD Idiota  contro Salvini ciabattoni  & Another PD-idiot against Salvini moron & \textsf{LE} & \textsf{LE} \\
        \hline
        $\surd$ & Ennesimo strafalcione geografico per il \#M5S: questa volta il vento del cambiamento sposta addirittura le regioni! Speriamo i fondi arrivino davvero in \#Molise, senza nulla togliere alle \#Marche.. & Yet another geographical blunder for the \#M5S: this time the wind of change even moves the regions! We hope the funds really arrive in \#Molise, without detracting anything from the \#Marche.. & \textsf{FdI} & \textsf{M5S} \\
        \hline
        $\surd$ & @virginiaraggi La compatisco. Fra un paio d'anni, allo scadere del suo mandato, lei finirà nell'oblio come merita. CasaPound sarà sempre al suo posto. & @virginiaraggi I sympathize with you. In a couple of years, at the end of your term, you will be forgotten as you deserve. CasaPound will always be in its place. & \textsf{CPI} & \textsf{CPI} \\
        \hline
        $\oslash$ & Le chiacchiere fanno i pidocchi, i maccheroni riempiono la pancia & The chatter makes the lice, the macaroni fill the belly & \textsf{FI} & \textsf{CPI} \\
        \hline
        $\oslash$ & In bocca al lupo alle ragazze e ai ragazzi della mia commissione. Domattina si parte \#notteprimadegliesami  & Good luck to the girls and boys of my commission. Tomorrow morning we begin \#nightbeforeexams & \textsf{PD} & \textsf{CPI} \\
    	\bottomrule
	\end{tabular}
	\end{adjustbox}
    \caption{Examples of tweets included in the enriched training-set of the party tweet classifier. Politically-relevant tweets are marked with $\surd$, irrelevant ones are marked with $\oslash$. The label initially assigned by $C$ to each example is reported in column $L_o$, while the label corrected through the usage of the minimal distance from a party account is reported in the $L_c$ column.}
	\label{tab:enriched_data_examples}
\end{table*}

In our preliminary experiments, we found that $Th = 0.5$ is a reasonable value to get both several thousands of new labeled examples and a large variety of new  tweets featuring substantial differences in their writing style, with respect to the typical tweets of an official party account. In detail, by applying the aforementioned process: (i) in step 2 we selected 1,462 users equally distributed among all parties, and (ii) at the end of step 3 we obtained an enrichment dataset composed of 8,753 new labeled tweets. As shown in Table~\ref{tab:enriched_data_examples}, this method is obviously not perfect but it ensures to obtain a plentiful variety of different examples that help to improve the precision and the generalizability of the enriched party classifier $C'$, in an unsupervised fashion.

\section{Predicting Political Leaning}
\label{sec:prediction}
\begin{figure*}[t]
    \centering
    \begin{subfigure}{.4\textwidth}%
        \includegraphics[width=\textwidth]{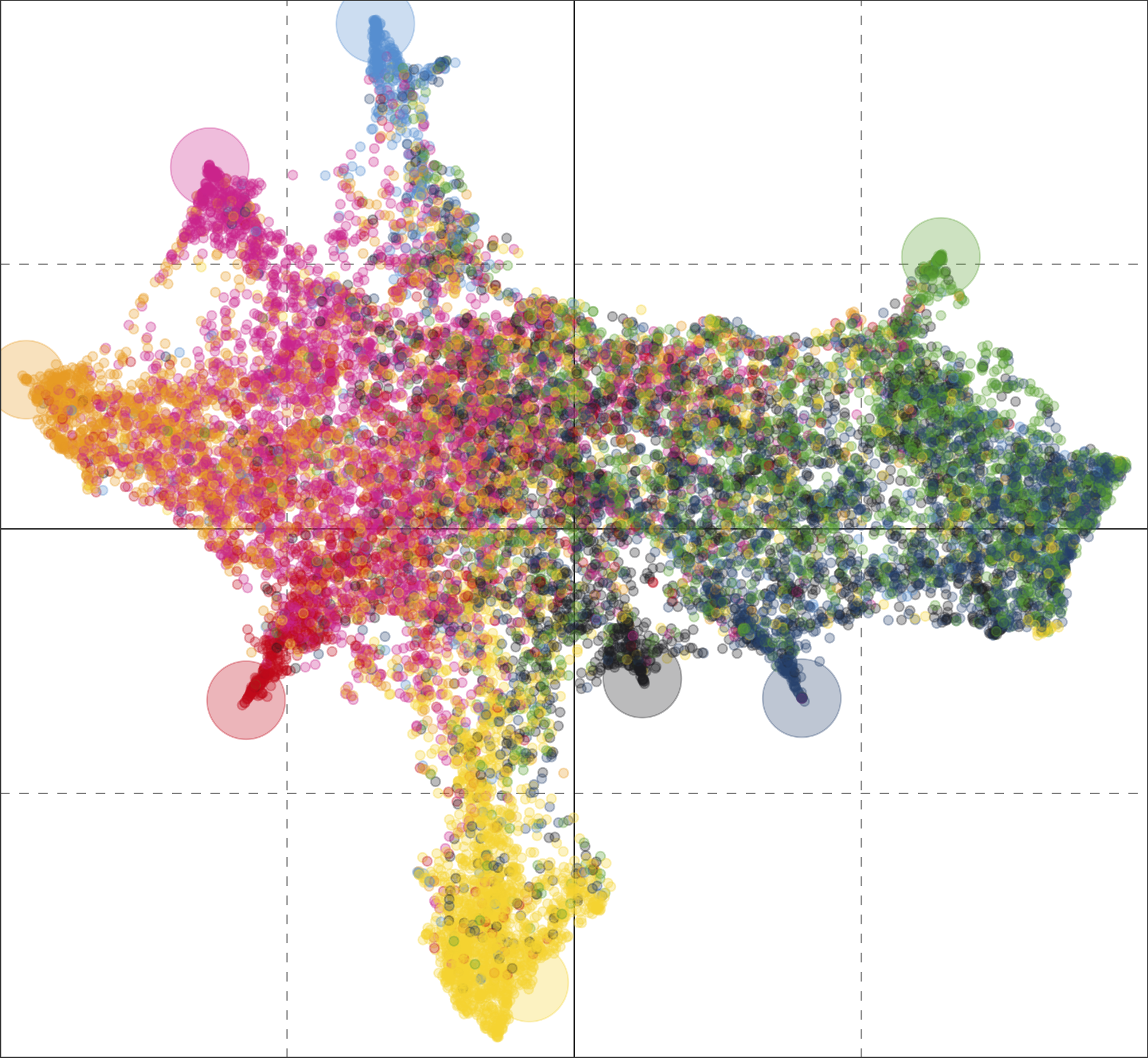}%
        \caption{Our technique.}
        \label{fig:party-projections-ours}
    \end{subfigure}%
    \hspace{.05\textwidth}%
    \begin{subfigure}{.472\textwidth}%
        \includegraphics[width=\textwidth]{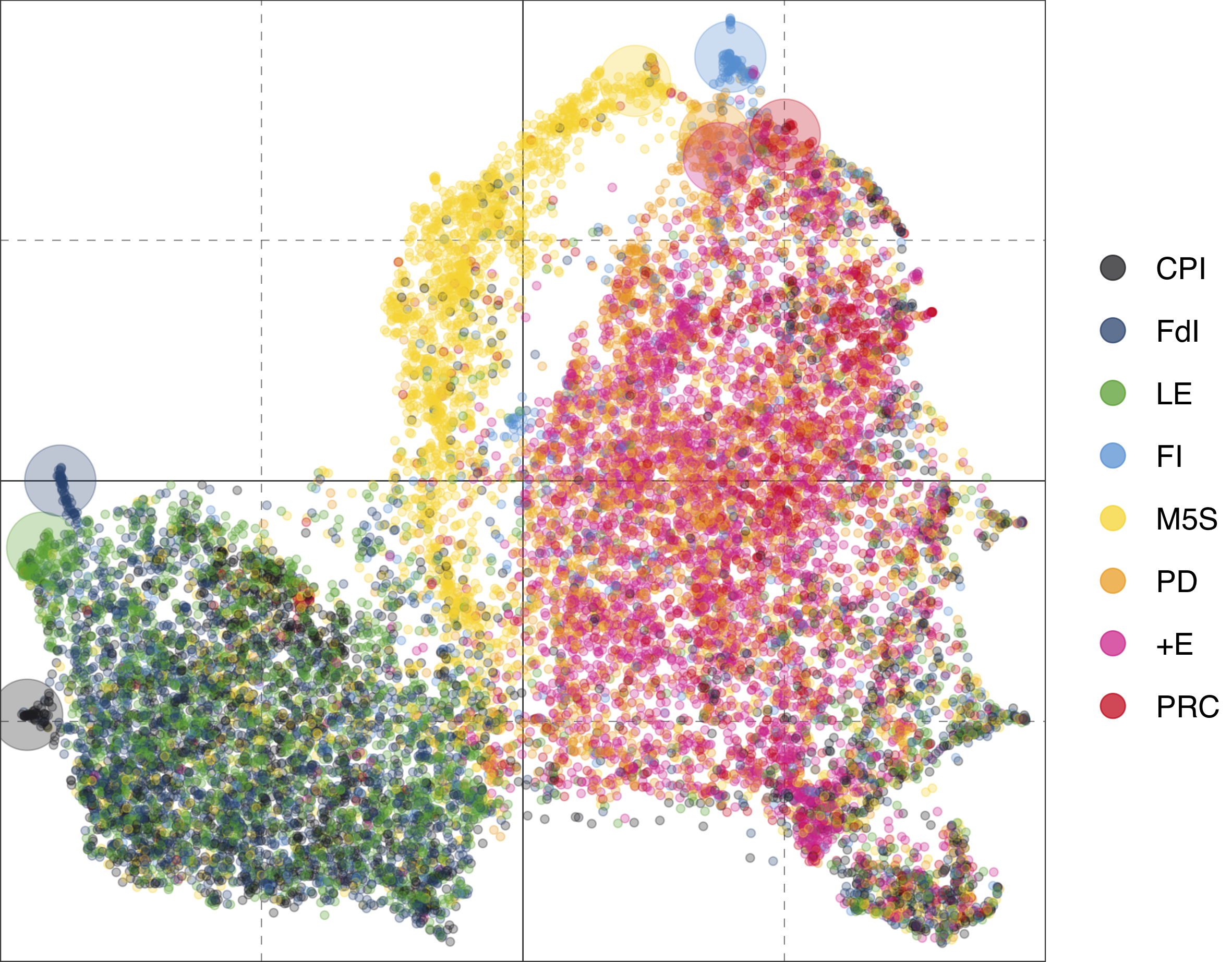}%
        \caption{Word2vec.}
        \label{fig:party-projections-w2v}
    \end{subfigure}%
    \caption{UMAP projections of the latent political ideology spaces learned by our proposed technique and via word2vec. Colors encode ground-truth party labels. Larger circles highlight the position of the official accounts for each considered party (i.e., our pivots).}
    \label{fig:party-projections}
\end{figure*}

\begin{figure*}[t]
    \centering
    \begin{subfigure}{.4\textwidth}%
        \includegraphics[width=\textwidth]{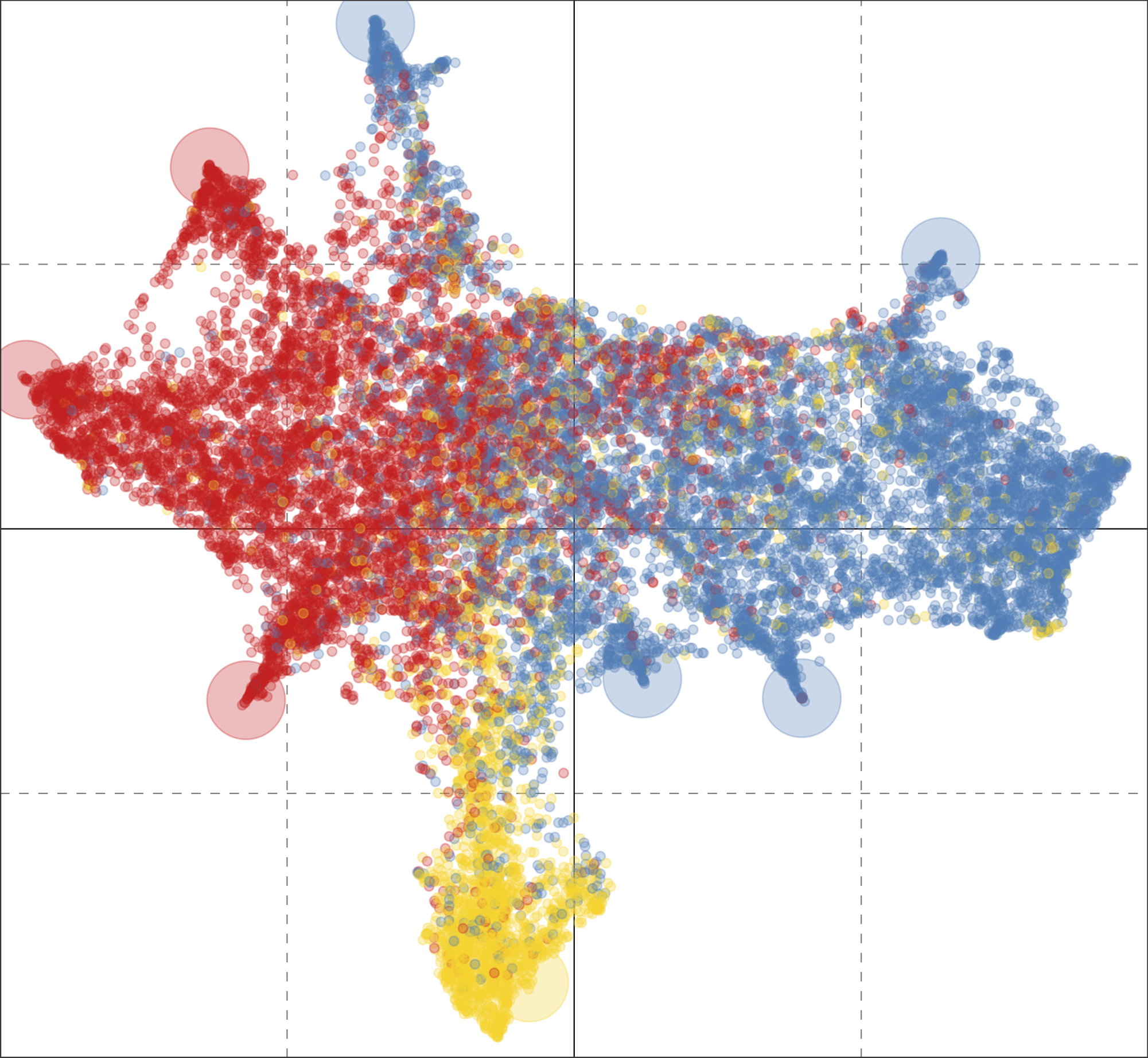}%
        \caption{Our technique.}
        \label{fig:pole-projections-ours}
    \end{subfigure}%
    \hspace{.05\textwidth}%
    \begin{subfigure}{.472\textwidth}%
        \includegraphics[width=\textwidth]{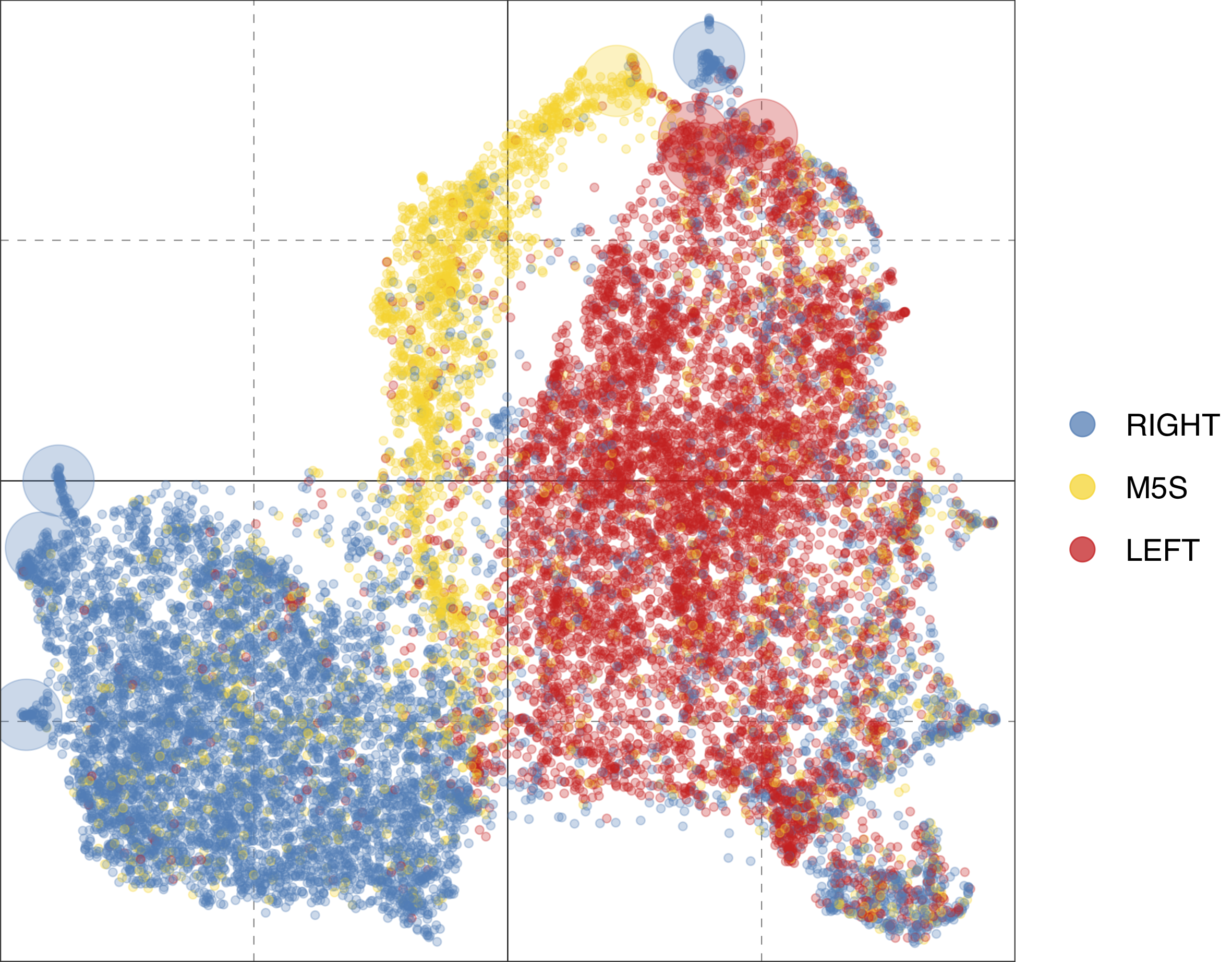}%
        \caption{Word2vec.}
        \label{fig:pole-projections-w2v}
    \end{subfigure}%
    \caption{UMAP projections of the latent political ideology spaces learned by our proposed technique and via word2vec. Colors encode ground-truth pole labels (right-leaning, left-leaning, and M5S). Larger circles highlight the position of the official accounts for each considered party (i.e., our pivots).}
    \label{fig:pole-projections}
\end{figure*}

By using the encoding scheme presented in Section \ref{sec:user-vector}, we are able to analyze the proposed method from a qualitative point of view and to map each user into a position within a shared latent political ideology space. This mapping is built directly onto user vectors with the aim of projecting users over a bidimensional geometric space in such a way to (i) minimize the distance between similar users having close political ideas and (ii) maximize the distance between users having different political opinions. To perform feature reduction and to map the latent user vectors in $\mathbb{R}^{40}$ to an equivalent space in $\mathbb{R}^{2}$, we leveraged UMAP with default parameters~\shortcite{mcinnes2018umap}. The mapping obtained from training data\footnote{Here, to better highlight data distribution in the political ideological space, we used training data because the amount of users is far bigger than those contained in test data and the distribution of points is practically the same (i.e., there is no drift between training and test data).} is shown in Figures~\ref{fig:party-projections-ours} and~\ref{fig:pole-projections-ours}, where users are respectively colored according to their party and pole labels. 

Regarding party projections, the first observation is that many users supporting a specific party are concentrated in the neighborhood of the party itself (indicated by large circle points). This is a quite strong indication that user feature representations provided by Equation~\eqref{eq:user-vector} properly describe the political stance of the parties. In addition, when users have enough political content in their timeline, the same method also allows to position them near to their preferred party. This first result is verified for all parties, and particularly so for the left-leaning ones and for the \textsf{M5S}. Regarding right-leaning parties (right-hand side of the figures), although this trend is confirmed, the situation is more fluid with users of \textsf{FI} clearly separated from the users of the other 3 right-leaning parties (\textsf{CPI}, \textsf{FdI} and \textsf{LE}). Indeed, supporters of the latter parties, in addition to forming clear clusters positioned around official party accounts, are spread over wide areas of the political ideology space, also creating regions where users of different parties are mixed together. This feature of our learned political ideology space is in agreement with the Italian political landscape, where these 3 far-right parties hold similar stances with respect to many political issues~\shortcite{pasquino2019state}, and with the opinions expressed by their electors. By analyzing Figure~\ref{fig:party-projections-ours}, it is also worth noting that even the central area of the ideology space contains a mixture of users belonging to different parties. Also this situation is expected and understandable, since it represents undecided users and users that hardly share any political content at all.

When considering pole projections shown in Figure \ref{fig:pole-projections-ours}, we can see that there is a clear separation between the three poles, with only the central region of the ideology space characterized by a physiological group of users whose political stance is not uniform, for the same motivations given before. Quite naturally, these qualitative results suggest that the fine-grained task (i.e., party prediction) represents a much more challenging problem than the coarse-grained one (i.e., pole prediction). This naturally results from the minimal differences between some of the considered parties.

For comparative purposes, in Figures \ref{fig:party-projections-w2v} and \ref{fig:pole-projections-w2v} we show the latent political ideology space obtained with a user encoding based on word2vec~\shortcite{mikolov2013efficient} instead of the one learned with our method. Word2vec is a very popular embeddings method that already demonstrated an excellent encoding power on many NLP tasks. In this case we used word2vec algorithm provided by gensim library\footnote{\url{https://radimrehurek.com/gensim/}} to build from scratch a custom model optimized for this specific context by learning the latent space directly from the used Twitter dataset. In order to build the new word2vec model, we applied a minimal step of preprocessing to raw textual data. In particular, we transformed texts into lowercase and removed all stopwords. With the resulting data, we built a word2vec model keeping only the most frequent 50,000 words. Each tweet is thus vectorized by computing the mean of the sum of vector embeddings of each word occurring in the text. The user vector is finally obtained by computing the mean of the tweet vectors extracted from the user's timeline. Differently from our method, the word2vec encoding seems unable to clearly separate the different political parties, as demonstrated by several regions of the ideology space featuring a mixture of users from different parties. Another major drawback of this approach is represented by the vicinity between the accounts of several different parties. While in Figure~\ref{fig:party-projections-ours} each pivot held a specific position in the ideology space, clearly separated from that of other parties, in Figure~\ref{fig:party-projections-w2v} several pivots end up laying in the same area of the ideology space, which inevitably hinders party separability and the prediction of users' political leaning. Regarding pole predictions, the situation improves. However, it does not reach the level of data separation obtained with our proposed technique. In summary, these findings suggest that word2vec encoding, in this particular context, is sub-optimal and not able to properly model the semantics of political ideologies of the different parties.

Based on the favorable properties of our learned latent political ideology space, the unsupervised prediction of user political leaning can be achieved by applying a clustering algorithm directly to the bidimensional projected user vectors. Without loss of generality, in this work we assume that we know the number of clusters we want to obtain at the end of clustering process (i.e., 8 clusters for the party prediction task and 3 clusters for pole prediction task). The steps needed by the clustering process are summarized the following:
\begin{enumerate}
    \item Projection of the users into a new bigger latent space based on the similarity of users. Each distinct user is seen as a separated feature in this new space and the feature vector of each user is generated by computing its pairwise distance to all the other users. This step was originally proposed by~\shortciteA{darwish2020unsupervised} and demonstrated to improve the subsequent clustering step.
    \item Feature reduction using UMAP to prevent the curse-of-dimensionality due to data sparseness~\shortcite{domingos2012few}.
    \item Feature standardization by subtracting the mean and scaling the features to unit variance.
    \item Data clustering using the KMeans, GaussianMixture, or MeanShift algorithms. This clustering step is based on the implementations provided by the \textit{sklearn} Python software package\footnote{\url{https://scikit-learn.org/stable/}}. 
\end{enumerate}

The first 3 steps of the above list are optional and can be used only in specific cases where they improve clustering accuracy. For the experiments reported in the next section, we followed the approach used by both~\shortciteA{darwish2020unsupervised} and~\shortciteA{di2018content}, and we evaluated different configurations on the validation partition of the dataset. Then, we used the best configuration obtained on the validation set to label users of the test set. The details about the specific configurations that we used are given in Section~\ref{sec:settings-comparisons}.

\section{Experiments and Results}
\label{sec:results}
Given the previously described method for predicting the political leaning of social media users, in this section we evaluate its predictions for test-set users of our dataset, with respect to the ground truth labels and to the predictions yielded by a number of baselines and other state-of-the-art techniques. Furthermore, we also validate our method by applying it to predict the political leaning of the Italian members of the European parliament (MEPs). Finally, we carry out a set of additional experiments to assess the sensitivity of our method to a number of relevant factors (e.g., distance from the pivots in the latent ideology space, number of tweets, number of retweets) and we provide a thorough analysis and discussion of our classification errors.

\subsection{Experimental Settings}
\label{sec:settings}
This section provides details on the evaluation settings and on the techniques used in our performance comparisons.

\subsubsection{Evaluation}
All techniques considered in our experiments are evaluated on two tasks. The aim of the first task is the prediction of fine-grained political leaning, which concerns with associating each user to its preferred political party. The second -- simpler -- task is the prediction of coarse-grained political leaning, where each user is assigned to a political pole (e.g., left-leaning, right-leaning, or leaning towards \textsf{M5S}). Similarly to previous work, we evaluate each task as a multiclass classification task. However, our experiments are considerably more challenging than those typically performed in previous works, due the larger number of involved classes. In fact, previous methods for predicting political leaning were typically evaluated in a binary classification setting (e.g., predicting left- \textit{vs} right-leaning users). Here instead, our fine-grained task encompasses 8 possible classes, while our coarse-grained task admits 3 classes. Given the moderate class imbalance for both the fine- and coarse-grained tasks, visible in Table~\ref{tab:politics}, for each evaluated method we report both the \textit{micro} and \textit{macro} versions of \textit{precision}, \textit{recall} and \textit{F1-measure}.

\subsubsection{Comparisons}
\label{sec:settings-comparisons}
In the upcoming sections, we report experimental results for several techniques, including different configurations of our present proposal, strong and weak baselines, and other state-of-the-art techniques. In the following, we briefly introduce each technique that we implemented and evaluated, starting from 3 interesting configurations of our proposal. Wherever meaningful, each technique is labeled by separately specifying the approach used for learning ideologies and that used for making predictions (i.e., {\small\textsf{ideologies + predictions}}).

\begin{description}
    \small{
    \item[Parties + clustering:] This method is based on the latent ideologies learned with our proposed unsupervised approach. For learning ideologies, we apply only the steps described in Sections~\ref{sec:party-classifier} and~\ref{sec:user-vector}, without the enrichment step introduced in Section~\ref{sec:enrichment}. Predictions are performed via clustering, as detailed in Section \ref{sec:prediction}, by applying step 1, step 2 with a feature reduction to 64 features, and step 4 using GaussianMixture with default parameters. These clustering settings are used for both prediction tasks.
    \item[Parties enriched + distance:] For this method we use unsupervised ideologies learned with all 3 steps described in Section~\ref{sec:ideology}, thus also including the enrichment step. Predictions are obtained by assigning each user to the party of the pivot the user is nearest to. This method represents a strong unsupervised baseline that leverages our learned ideologies, combined with a simple prediction strategy.
    \item[Parties enriched + clustering:] This is our most complete method. It is fully unsupervised, it leverages all steps for learning latent ideologies as well as clustering for obtaining predictions. The clustering process for the party prediction task is performed by applying only step 3 and step 4 using KMeans as the clustering algorithm. For the pole prediction task, we used instead step 1, step 2 with a feature reduction to 64 features, and step 4 using the KMeans algorithm.}
\end{description}

In addition to our 3 unsupervised contributions reported above, we also experimented with a number of baselines and other approaches, which we briefly describe in the following.

\begin{description}
    \small{
    \item[Random classifier:] Simple unsupervised baseline that outputs random predictions.
    \item[Majority classifier:] Simple supervised baseline that always outputs the majority class.
    \item[Word2vec + clustering:] Unsupervised method where latent ideologies are learned by leveraging word2vec embeddings, while predictions on both tasks are obtained via clustering applying only step 4 with the KMeans algorithm.
    \item[Retweets + clustering:] This method implements the state-of-the-art unsupervised technique proposed by~\shortciteA{darwish2020unsupervised}. It first learns user representations based on user retweets, then it obtains predictions on both tasks via clustering by applying step 1 and step 4 implemented with the MeanShift clustering algorithm.
    \item[Supervised enriched + clustering:] This method is similar to the \textsf{parties enriched + clustering} one, with the exception of how the enrichment step is carried out. To this end, this method feeds back into the tweet classifier those tweets for which the classifier outputted wrong predictions despite having a high confidence. Given that this method exploits ground-truth labels for the enrichment step and clustering for obtaining predictions, it is classified as semi-supervised. The clustering process is performed by applying step 1, step 2 and step 4. On the party prediction task, we reduced features to 64 dimensions and we used KMeans as the clustering algorithm. Differently, on the pole prediction task, we reduced features to 2 dimensions and we applied the GaussianMixture algorithm with default parameters.
    \item[Parties enriched + SVC:] This overall semi-supervised method leverages all our proposed steps for learning unsupervised latent ideologies, including enrichment. Then, predictions are obtained by training a supervised SVC classifier.
    \item[Supervised enriched + SVC:] Here we exploit our supervised enriched ideologies in conjunction with an SVC classifier. Both steps used in this method are thus supervised.
    \item[Word2vec + SVC:] Ideologies used in this semi-supervised method are obtained via word2vec embeddings, while predictions are yielded by an SVC classifier.}
\end{description}

\subsection{Results}
\label{sec:res}
In the remainder of this section we present and discuss experimental results for the 2 considered tasks: coarse- and fine-grained prediction of political leaning. We first compare results of our method to those of the several others that we evaluated. While discussing such comparisons, we particularly focus on the differences in performance between our best proposal and the technique introduced by~\shortciteA{darwish2020unsupervised}, which is considered the state-of-the-art for unsupervised prediction of political leaning. Then, we measure and discuss the performance gap between unsupervised approaches with respect to semi-supervised and supervised ones. Finally, we investigate the impact of retweets and distance from pivots, for predicting the political leaning of social media users.

\subsubsection{Prediction of Political Leaning: Unsupervised Approaches}
\label{sec:res-unsupervised}
We begin by evaluating the performance of unsupervised techniques on the fine-grained prediction task. This is the core contribution of our work and results of this evaluation and comparison are shown in Table~\ref{tab:party_unsupervised_results}. Our 3 contributions are compared to a \textsf{random classifier}, to the technique proposed by~\shortciteA{darwish2020unsupervised} and to an approach based on \textsf{word2vec + clustering}.

\begin{table*}[t]
    \footnotesize
    \centering
    \begin{adjustbox}{max width=\textwidth}
	\begin{tabular}{p{0.05em}llcrrrcrrr}
        \toprule
        & \multicolumn{2}{c}{\textbf{method}} && \multicolumn{3}{c}{\textbf{macro}} && \multicolumn{3}{c}{\textbf{micro}} \\
        \cmidrule{2-3}\cmidrule{5-7}\cmidrule{9-11}
        & \textit{ideologies} & \textit{predictions} && \textit{precision} & \textit{recall} & \textit{F1} && \textit{precision} & \textit{recall} & \textit{F1} \\
    	\midrule
    	& -- & \textsf{random classifier} && 0.124 & 0.125 & 0.120 && 0.143 & 0.126 & 0.131 \\
    	& \textsf{word2vec} & \textsf{clustering} && 0.128 & 0.111 & 0.106 && 0.171 & 0.139 & 0.142 \\
    	{\scriptsize$\ddag$}& \textsf{retweets} & \textsf{clustering} && 0.370 & 0.293 & 0.301 && 0.420 & 0.346 & 0.354 \\
    	\midrule
    	& \multicolumn{2}{c}{\textit{our contributions}} \\ [0.6em]
    	& \textsf{parties} & \textsf{clustering} && 0.390 & 0.344 & 0.342 &&  0.443 & 0.354 & 0.372 \\
    	& \textsf{parties enriched} & \textsf{distance} && 0.419 & 0.370 & 0.324 && 0.489 & 0.339 & 0.330 \\
    	& \textsf{parties enriched} & \textsf{clustering} && \textbf{0.472} & \textbf{0.434} & \textbf{0.426} && \textbf{0.517} & \textbf{0.421} & \textbf{0.426} \\
    	\bottomrule
    	\multicolumn{11}{l}{\scriptsize$\ddag$:~\shortcite{darwish2020unsupervised}}
	\end{tabular}
	\end{adjustbox}
    \caption{Performance comparison of \textit{unsupervised} methods for fine-grained (party) prediction of political leaning. The best result for each evaluation metric is shown in bold font.}
	\label{tab:party_unsupervised_results}
\end{table*}

All results reported in Table~\ref{tab:party_unsupervised_results} are moderate, at best. This shows the difficulty of the fine-grained task. Among the evaluated techniques, our \textsf{parties enriched + clustering} method achieves the best results in each evaluation metric, with \textit{micro} and \textit{macro F1} $= 0.426$. This is our most complete proposal that takes full advantage of all the steps described in Sections~\ref{sec:ideology} and~\ref{sec:prediction}. The second-best method is \textsf{parties + clustering}, with \textit{micro F1} $= 0.372$ and \textit{macro F1} $= 0.342$. The difference in performance between these 2 methods is solely due to the enrichment step, that we introduced in Section~\ref{sec:enrichment}, and that allows learning better user representations as demonstrated by these results. The state-of-the-art technique by~\shortciteA{darwish2020unsupervised} achieves the third-best results, confirming its overall value.
Interestingly, the \textsf{parties enriched + distance} strong baseline achieves performances that are only slightly worse than those of the previous methods. This further demonstrates the informativeness of the latent user ideologies that we learned. Contrarily, both the \textsf{word2vec + clustering} and the simple \textsf{random classifier} baseline obtain unsatisfactory performances, with \textit{micro F1} $= 0.142$ and $0.131$, respectively. This result is particularly interesting for the \textsf{word2vec + clustering} approach. In fact, it suggests that the user representations learned by word2vec in this context, are not suitable for a prediction step via clustering.

\begin{table*}[t]
    \footnotesize
    \centering
    \begin{adjustbox}{max width=\textwidth}
	\begin{tabular}{p{0.05em}llcrrrcrrr}
        \toprule
        & \multicolumn{2}{c}{\textbf{method}} && \multicolumn{3}{c}{\textbf{macro}} && \multicolumn{3}{c}{\textbf{micro}} \\
        \cmidrule{2-3}\cmidrule{5-7}\cmidrule{9-11}
        & \textit{ideologies} & \textit{predictions} && \textit{precision} & \textit{recall} & \textit{F1} && \textit{precision} & \textit{recall} & \textit{F1} \\
    	\midrule
    	& -- & \textsf{random classifier} && 0.323 & 0.321 & 0.310 && 0.370 & 0.324 & 0.339 \\
    	& \textsf{word2vec} & \textsf{clustering} && 0.429 & 0.426 & 0.415 && 0.491 & 0.480 & 0.471 \\
    	{\scriptsize$\ddag$}& \textsf{retweets} & \textsf{clustering} && \textbf{0.804} & 0.657 & 0.688 && 0.758 & 0.719 & 0.710 \\
    	\midrule
    	& \multicolumn{2}{c}{\textit{our contributions}} \\ [0.6em]
    	& \textsf{parties} & \textsf{clustering} && 0.599 & 0.587 & 0.586 && 0.665 & 0.633 & 0.640 \\
    	& \textsf{parties enriched} & \textsf{distance} && 0.758 & 0.575 & 0.569 && 0.728 & 0.698 & 0.659 \\
    	& \textsf{parties enriched} & \textsf{clustering} && 0.751 & \textbf{0.752} & \textbf{0.750} && \textbf{0.776} & \textbf{0.772} & \textbf{0.772} \\
    	\bottomrule
    	\multicolumn{11}{l}{\scriptsize$\ddag$:~\shortcite{darwish2020unsupervised}}
	\end{tabular}
	\end{adjustbox}
	\caption{Performance comparison of \textit{unsupervised} methods for coarse-grained (pole) prediction of political leaning. The best result for each evaluation metric is shown in bold font.}
	\label{tab:pole_unsupervised_results}
\end{table*}

Table~\ref{tab:pole_unsupervised_results} shows the evaluation results of the same methods for the coarse-grained task. Difficulty-wise, this task is similar to those tackled in previous works~\shortcite{barbera2015birds,kulshrestha2017quantifying,di2018content,darwish2020unsupervised}. As such, these results are comparable to those reported in existing literature. In particular, all methods greatly improve and the best ones obtain rather good performances. As for the fine-grained task, the \textsf{parties enriched + clustering} method achieves the best results, with a balanced and promising \textit{micro F1} $= 0.772$ and \textit{macro F1} $= 0.750$. Also the method by~\shortciteA{darwish2020unsupervised} greatly improves, reaching the second-best overall result with \textit{micro F1} $= 0.710$ and \textit{macro F1} $= 0.688$, and the best \textit{macro precision}. The 2 other techniques based on our approach obtain comparable results, with \textit{micro F1} $\approx 0.65$ and \textit{macro F1} $\approx 0.57$. Finally, \textsf{word2vec + clustering} and the \textsf{random classifier} again obtain markedly worse results, thus confirming the underwhelming performance already measured for the fine-grained task.

\begin{figure*}[t]
    \centering
    \begin{subfigure}{.5\textwidth}%
        \includegraphics[width=\textwidth]{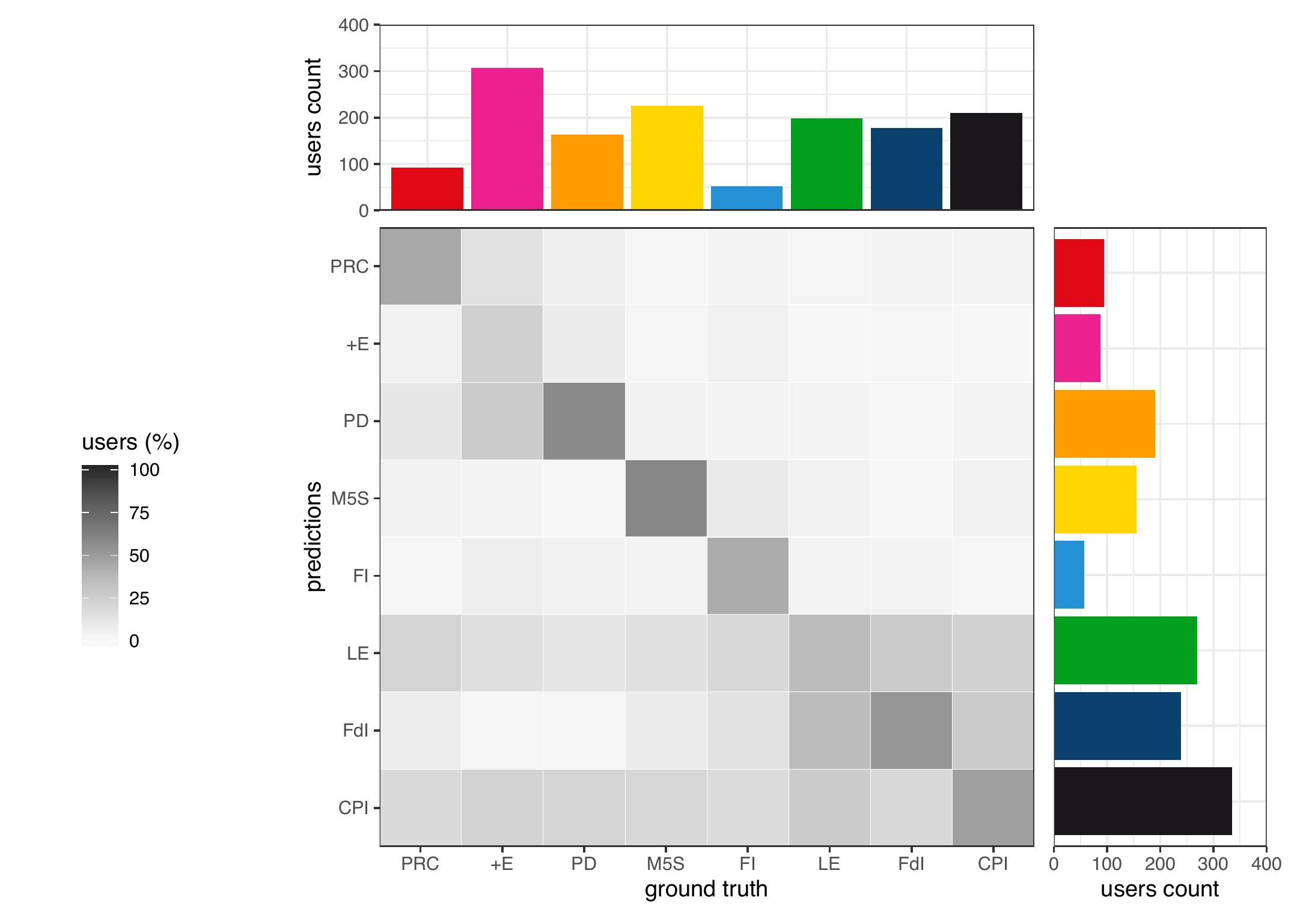}
        \caption{Our technique.}
        \label{fig:party-confusion-ours}
    \end{subfigure}%
    \begin{subfigure}{.5\textwidth}%
        \includegraphics[width=\textwidth]{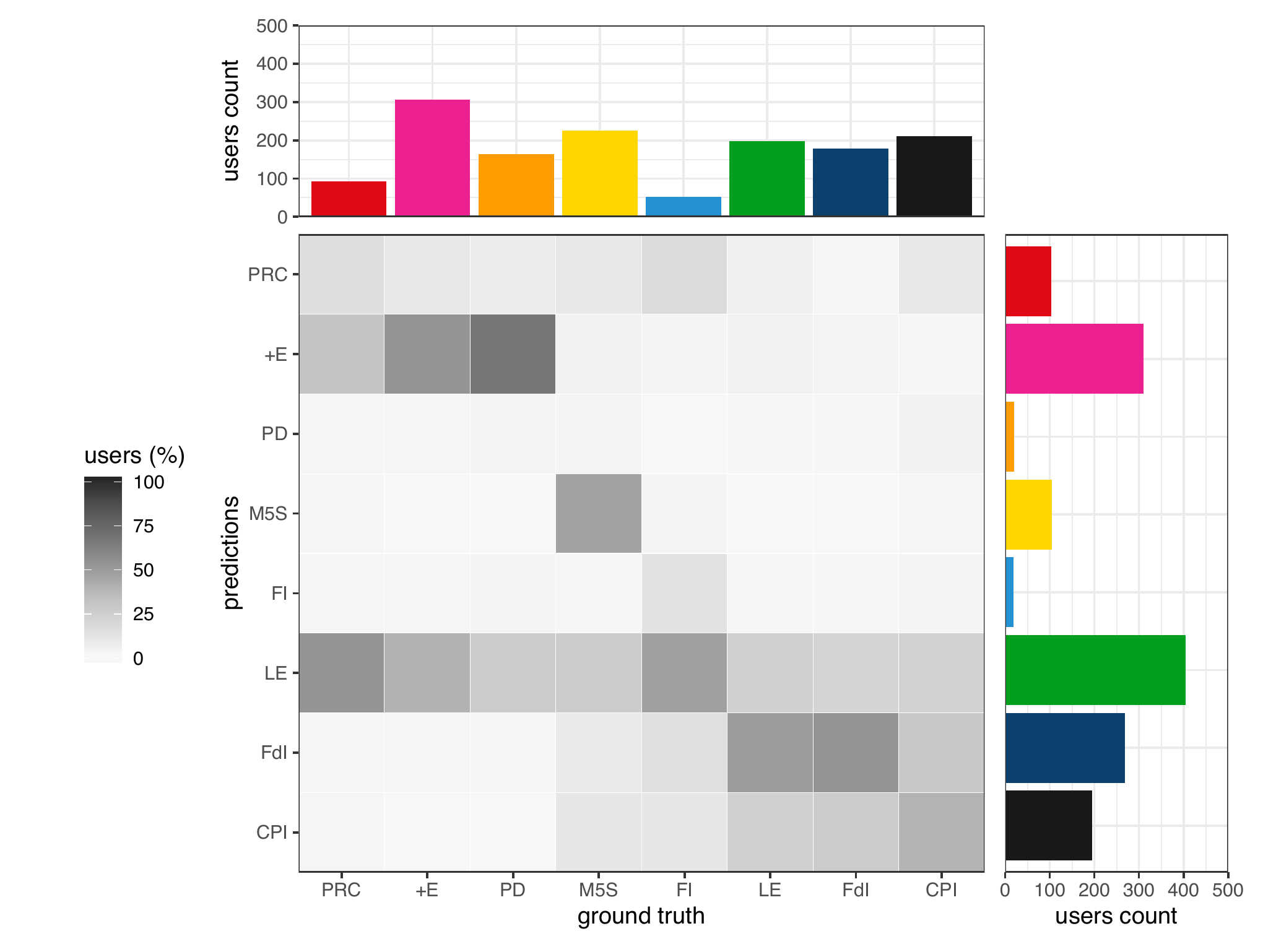}
        \caption{Darwish \textit{et al}.}
        \label{fig:party-confusion-darwish}
    \end{subfigure}%
    \caption{Comparison of the confusion matrices, with marginal distributions, for fine-grained (party) predictions between our proposed technique and the unsupervised method by~\shortciteA{darwish2020unsupervised}. Correct predictions lay on the matrix diagonal.}
    \label{fig:party-confusion}
\end{figure*}

The large improvement measured by both our \textsf{parties enriched + clustering} method and the technique by~\shortciteA{darwish2020unsupervised} between Tables~\ref{tab:party_unsupervised_results} and~\ref{tab:pole_unsupervised_results}, demonstrate that many of the mistakes made by these techniques in the fine-grained task consisted in misclassifying a user of a given party to a different party of the same pole, rather than to a party to the opposite of the political spectrum. This is expected and is due to the difficulty of the fine-grained task. Furthermore, it explains why the same techniques obtain strikingly better results when evaluated for the prediction of poles instead of parties. Figure~\ref{fig:party-confusion} helps to clarify this point. It shows the fine-grained confusion matrices of the 2 techniques, together with the marginal distributions of both ground-truth and predicted labels. First of all, the figure clearly highlights that more data points lay on the matrix diagonal in Figure~\ref{fig:party-confusion-ours} compared to Figure~\ref{fig:party-confusion-darwish}. This visually explains the better overall performance of our technique with regards to that by~\shortciteA{darwish2020unsupervised}. In addition, it shows the existence of two darker-colored $3\times3$ squares laying in the bottom-right and top-left corner of Figure~\ref{fig:party-confusion-ours}. To a lower extent, the same also occurs in  Figure~\ref{fig:party-confusion-darwish}. These regions of the confusion matrices allow to visualize the mistakes that we mentioned earlier -- that is, wrong party classifications that become correct predictions when techniques are evaluated pole-wise. The fact that these regions are more visible in Figure~\ref{fig:party-confusion-ours} than in Figure~\ref{fig:party-confusion-darwish} explains the better results in Table~\ref{tab:pole_unsupervised_results} of our technique with respect to~\shortciteA{darwish2020unsupervised}, especially regarding the \textit{macro F1} metric where we achieve $0.750$ \textit{vs} $0.688$. Figure~\ref{fig:party-confusion} also shows that our technique is particularly good at predicting center-leaning parties, such as \textsf{PD}, \textsf{M5S} and \textsf{FI}, for which we obtain almost flawless predictions. Contrarily, we have more difficulties in predicting far-right and far-left parties. Moreover, both techniques exhibit a bias towards overestimating right-leaning parties. On top of that,~\shortciteauthor{darwish2020unsupervised} also overestimate \textsf{+E} and almost completely neglect \textsf{PD} and \textsf{FI}.

\begin{figure*}[t]
    \centering
    \begin{subfigure}{.24\textwidth}%
        \includegraphics[width=\textwidth]{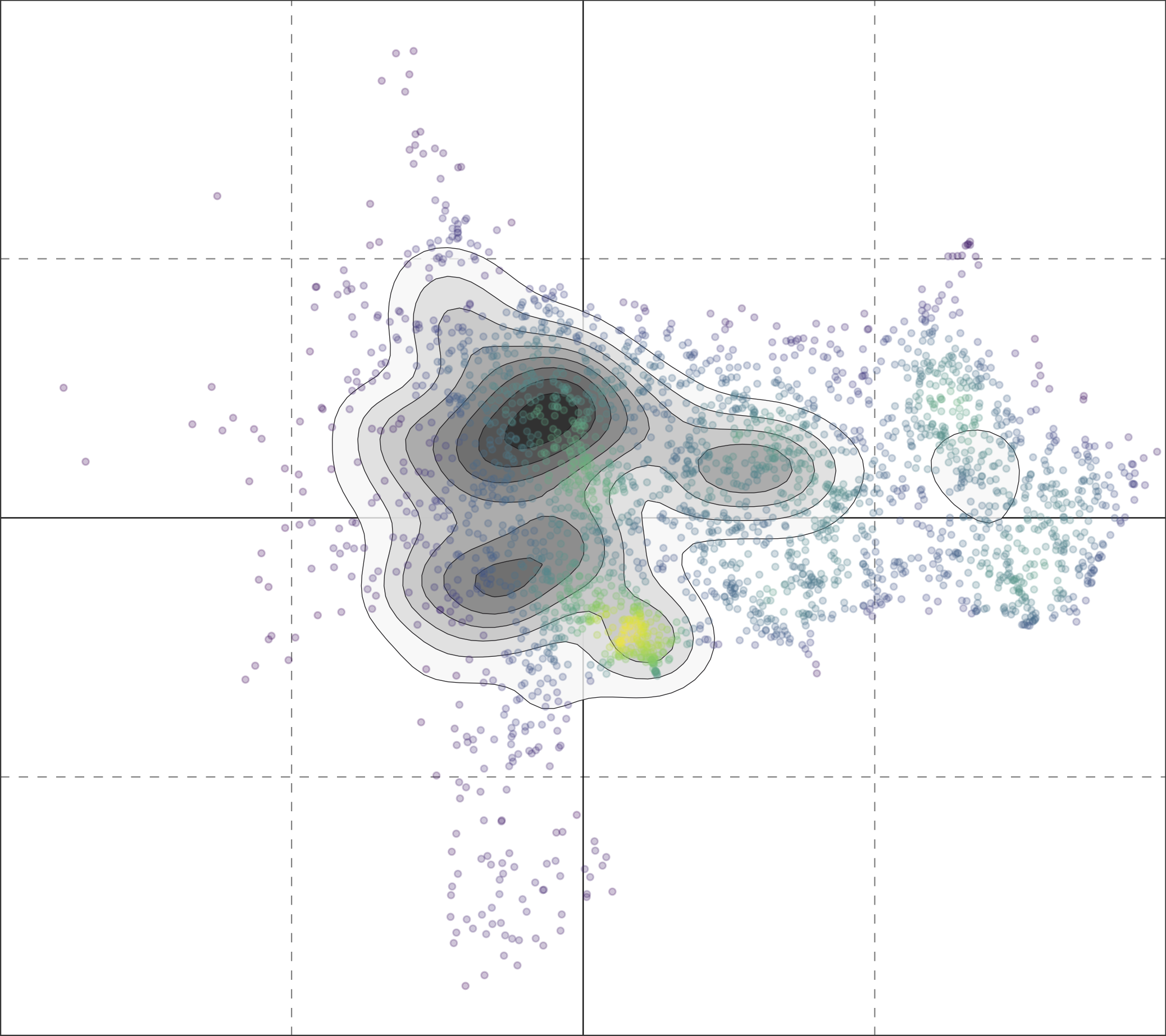}
        \caption{\tikz\draw[black, fill=CPI] (0,0) circle (.85ex); \textsf{CPI}.}
        \label{fig:party-densities-cpi}
    \end{subfigure}%
    \hspace{.013\textwidth}%
    \begin{subfigure}{.24\textwidth}%
        \includegraphics[width=\textwidth]{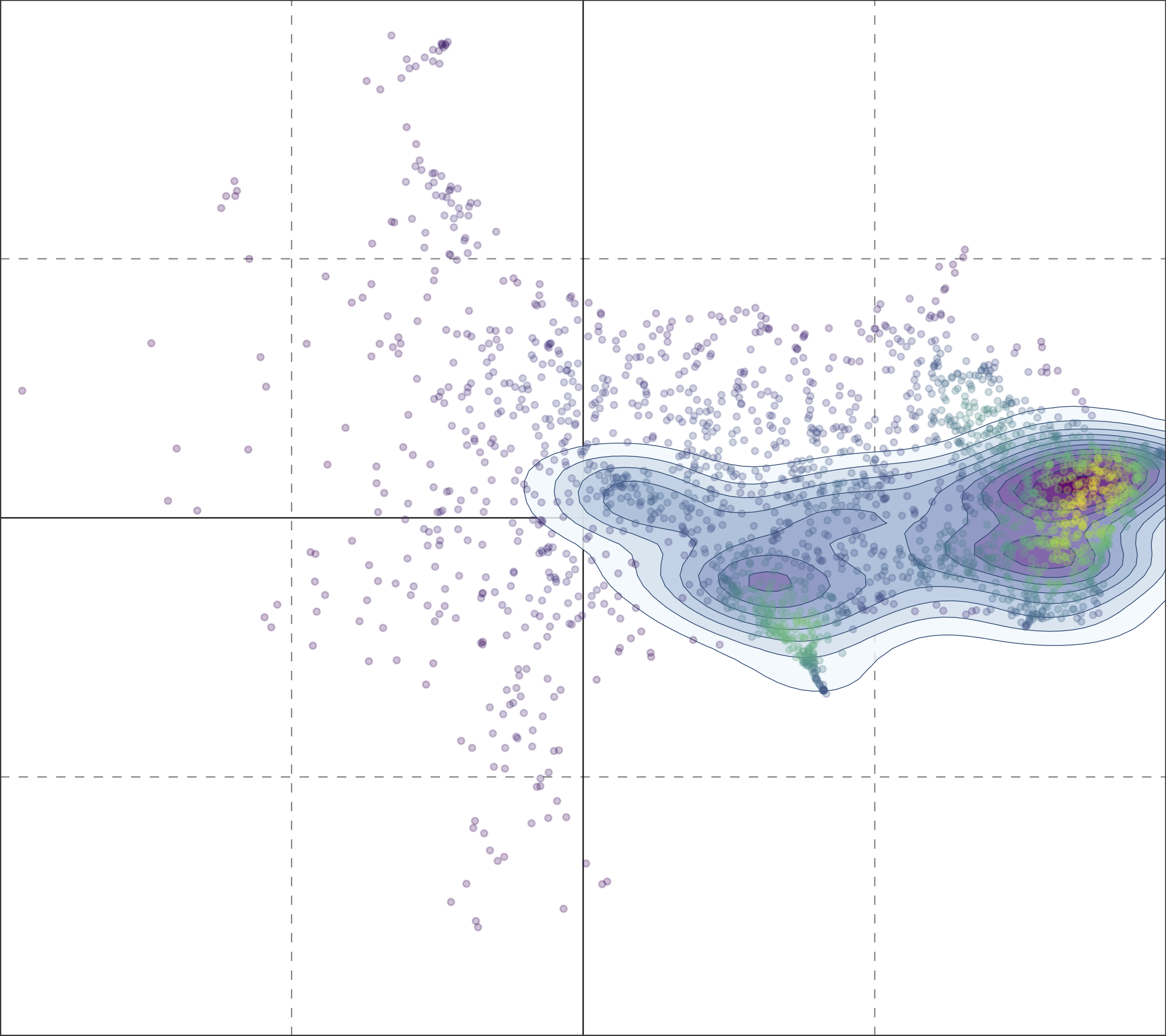}
        \caption{\tikz\draw[black, fill=FdI] (0,0) circle (.85ex); \textsf{FdI}.}
        \label{fig:party-densities-fdi}
    \end{subfigure}%
    \hspace{.013\textwidth}%
    \begin{subfigure}{.24\textwidth}%
        \includegraphics[width=\textwidth]{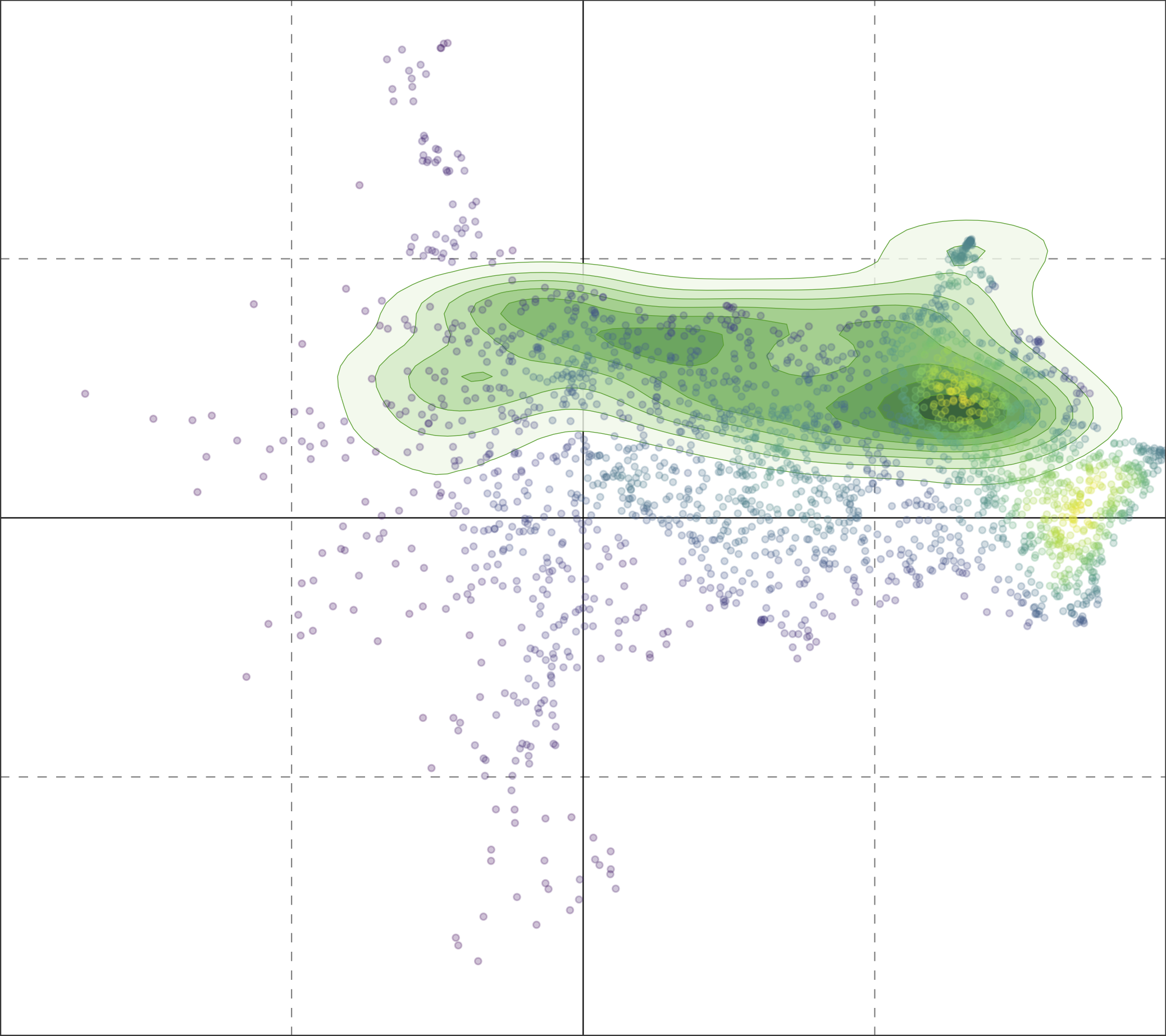}
        \caption{\tikz\draw[black, fill=LE] (0,0) circle (.85ex); \textsf{LE}.}
        \label{fig:party-densities-le}
    \end{subfigure}%
    \hspace{.013\textwidth}%
    \begin{subfigure}{.24\textwidth}%
        \includegraphics[width=\textwidth]{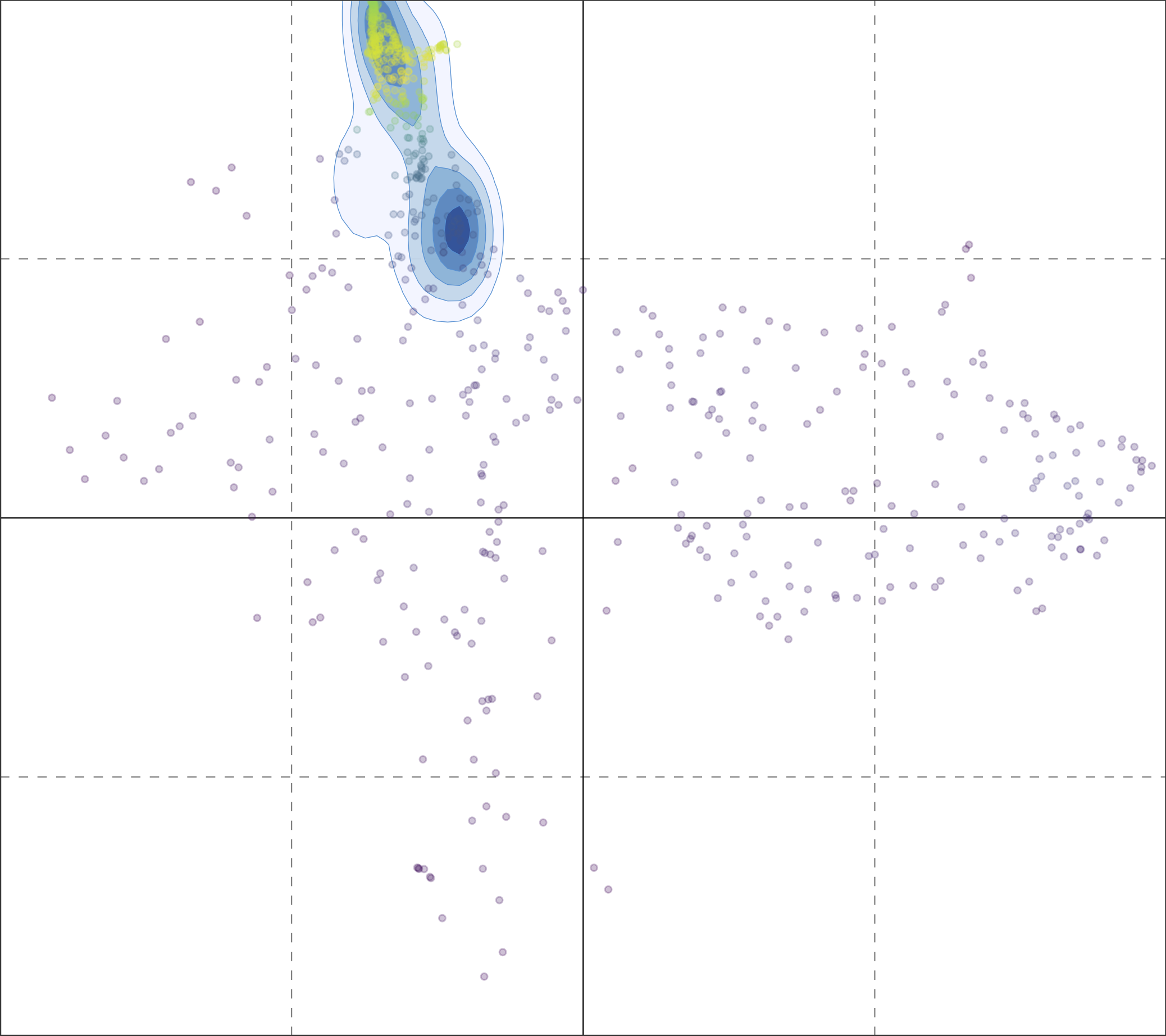}
        \caption{\tikz\draw[black, fill=FI] (0,0) circle (.85ex); \textsf{FI}.}
        \label{fig:party-densities-fi}
    \end{subfigure}\\
    \medskip
    \begin{subfigure}{.24\textwidth}%
        \includegraphics[width=\textwidth]{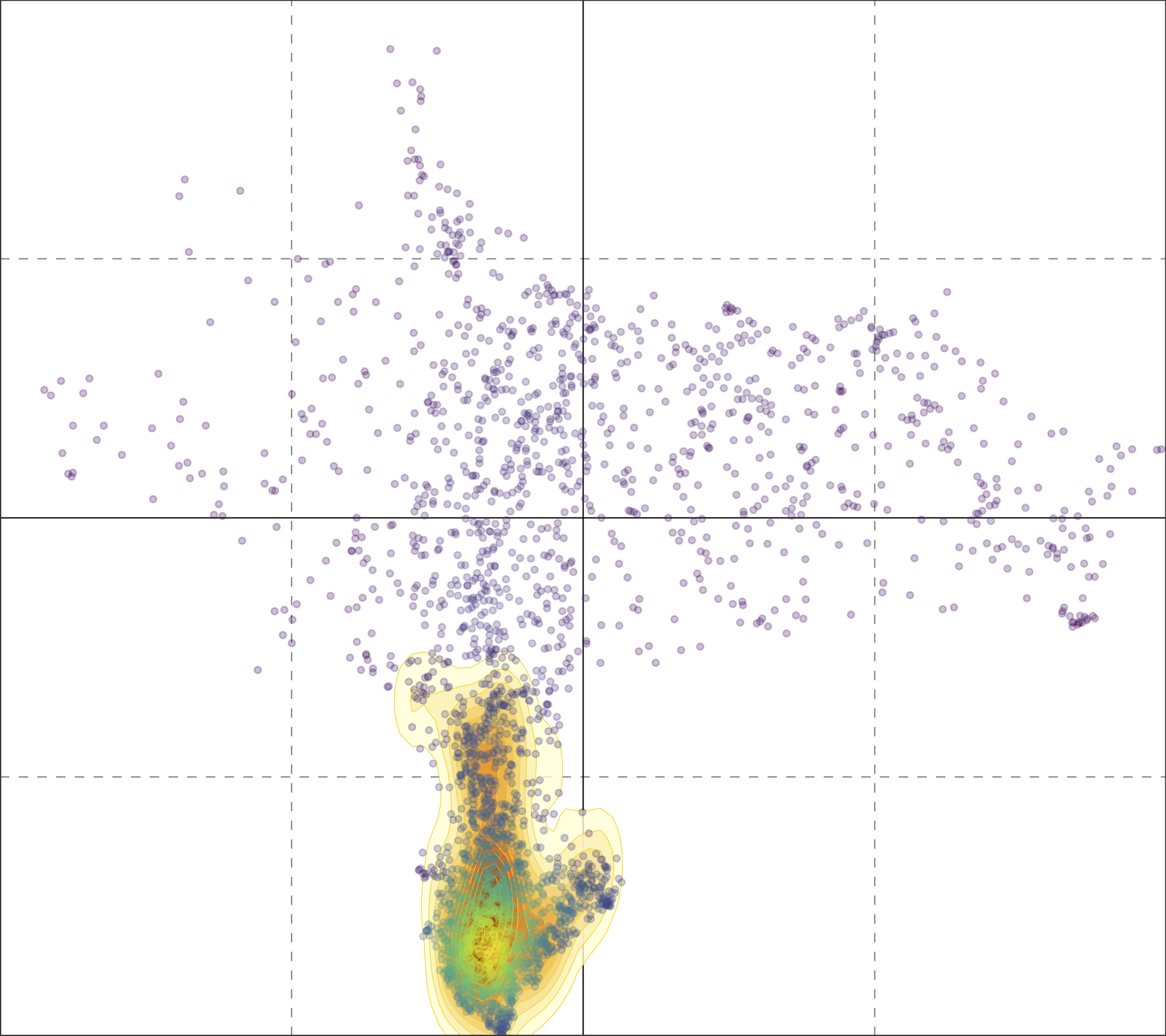}
        \caption{\tikz\draw[black, fill=M5S] (0,0) circle (.85ex); \textsf{M5S}.}
        \label{fig:party-densities-m5s}
    \end{subfigure}%
    \hspace{.013\textwidth}%
    \begin{subfigure}{.24\textwidth}%
        \includegraphics[width=\textwidth]{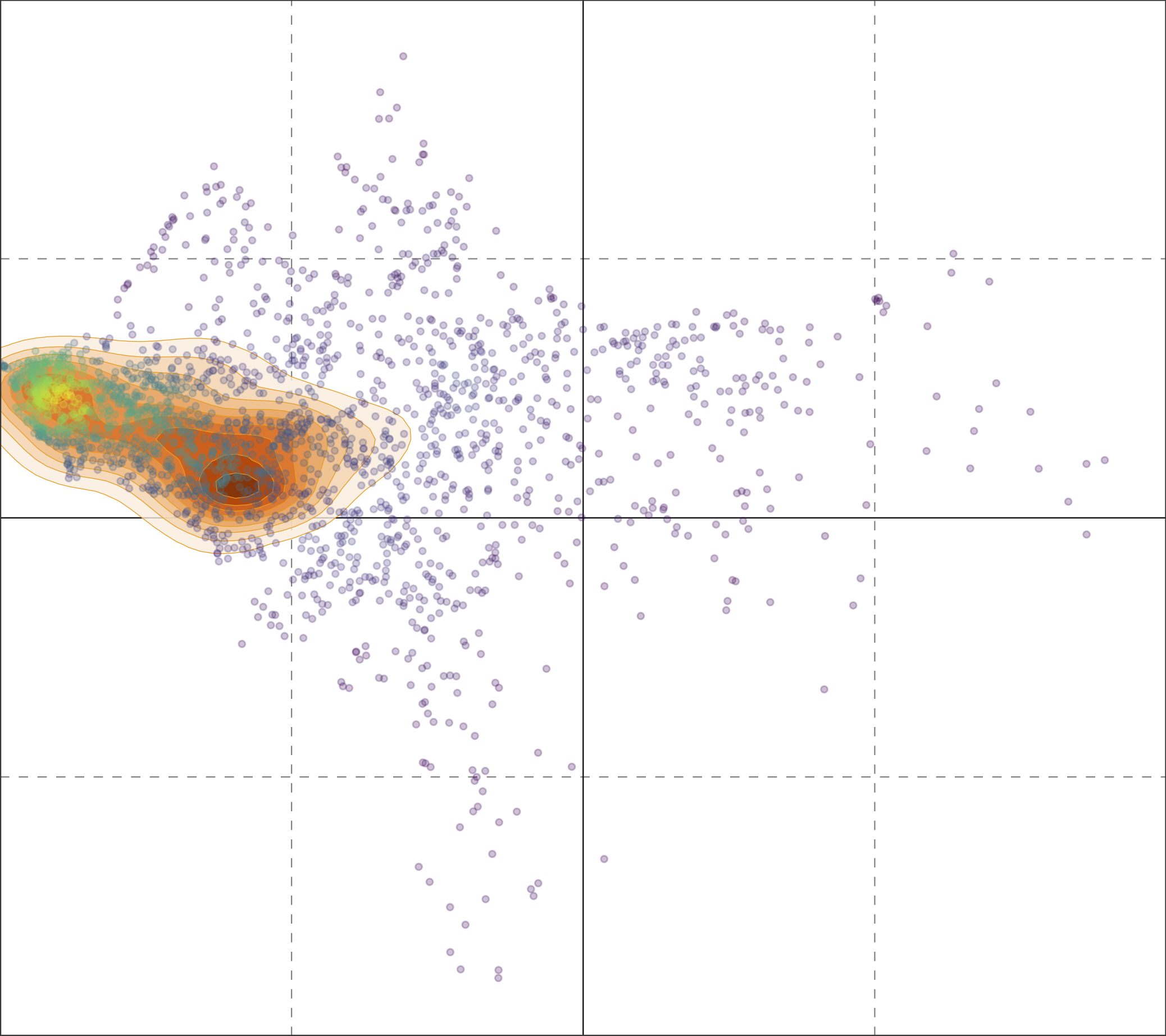}
        \caption{\tikz\draw[black, fill=PD] (0,0) circle (.85ex); \textsf{PD}.}
        \label{fig:party-densities-pd}
    \end{subfigure}%
    \hspace{.013\textwidth}%
    \begin{subfigure}{.24\textwidth}%
        \includegraphics[width=\textwidth]{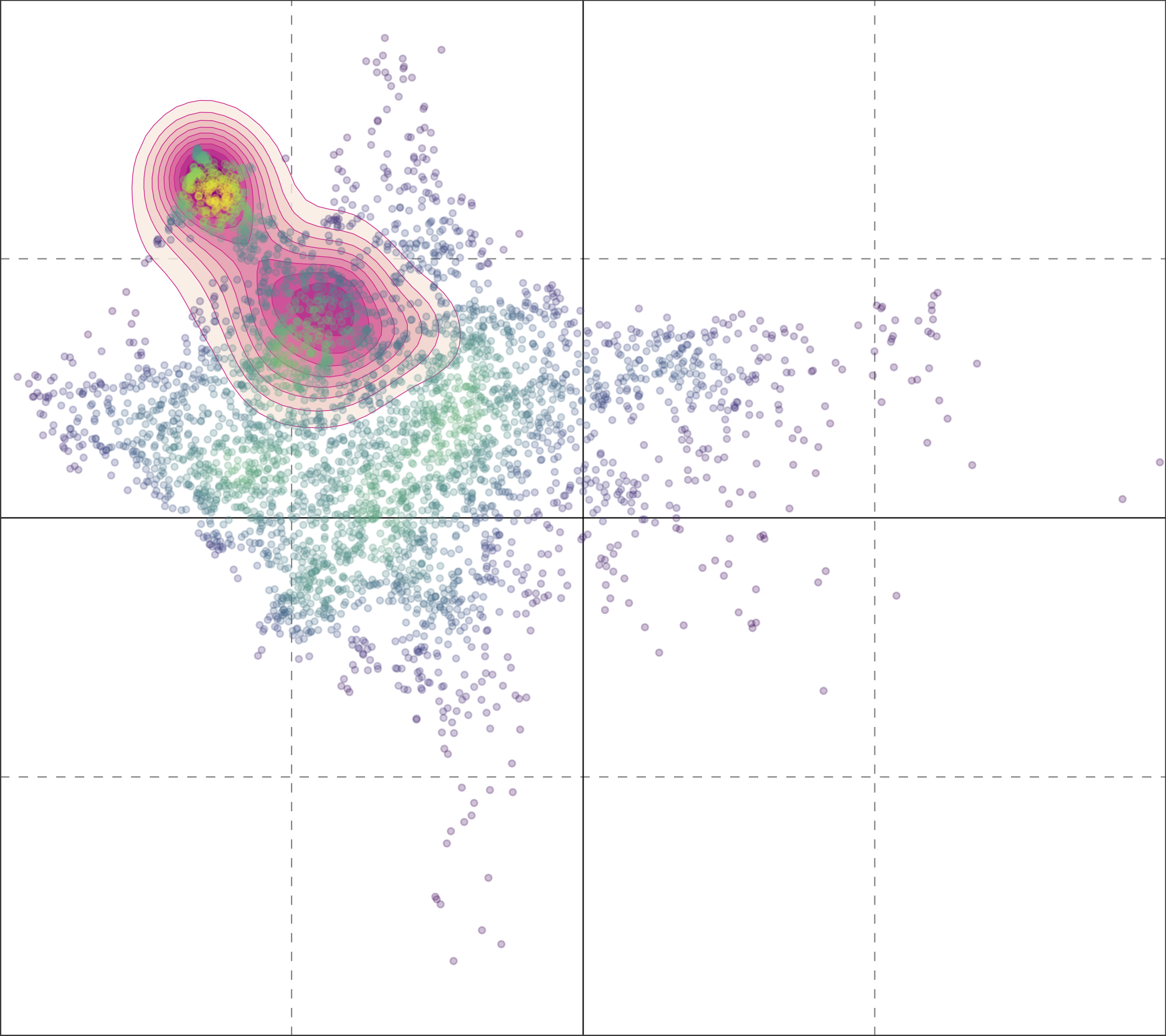}
        \caption{\tikz\draw[black, fill=+E] (0,0) circle (.85ex); \textsf{+E}.}
        \label{fig:party-densities-e}
    \end{subfigure}%
    \hspace{.013\textwidth}%
    \begin{subfigure}{.24\textwidth}%
        \includegraphics[width=\textwidth]{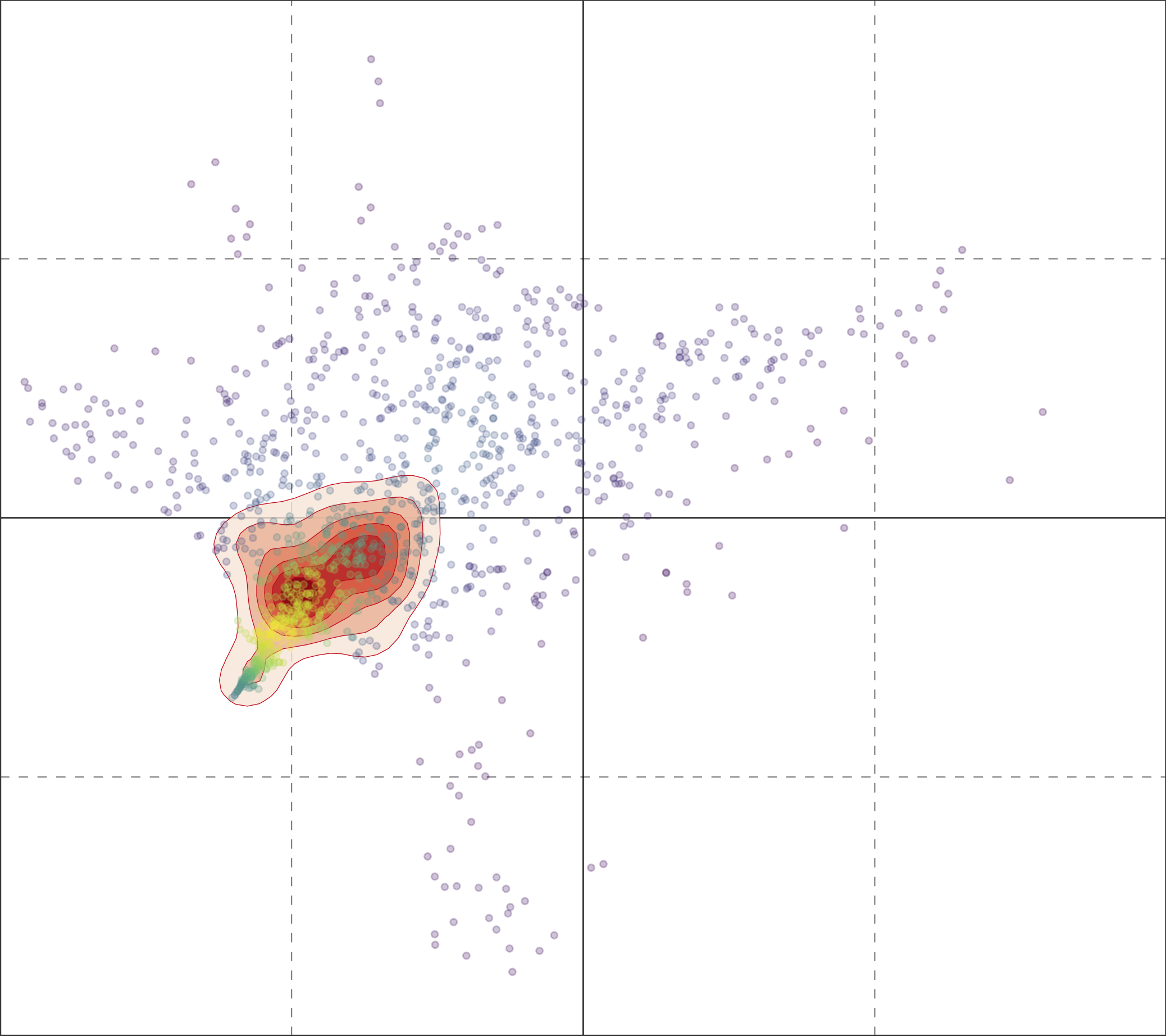}
        \caption{\tikz\draw[black, fill=PRC] (0,0) circle (.85ex); \textsf{PRC}.}
        \label{fig:party-densities-prc}
    \end{subfigure}%
    \caption{Comparison between the distribution of ground-truth users and of our predictions, within the latent ideology space. For each party, the distribution of ground-truth users is shown as a density-colored scatter plot. The distribution of our predictions is shown with contour lines.}
    \label{fig:party-densities}
\end{figure*}

Overall, results presented in Tables~\ref{tab:party_unsupervised_results} and~\ref{tab:pole_unsupervised_results} and in Figure~\ref{fig:party-confusion} demonstrate that it is very challenging to distinguish between the different parties that lay on the same side of the political spectrum. In order to provide an even more detailed breakdown of our party predictions, in Figure~\ref{fig:party-densities} we show ground-truth and prediction densities, for each party. Specifically in each subfigure, the scatter plot distribution shows where ground-truth users of a given party are positioned within the shared ideology space. Overlaid, the contour lines show the distribution of the test-set users predicted by our technique for that party. In other words, the maps of our ideology space shown in Figure~\ref{fig:party-densities} somewhat resemble the saliency maps used in computer vision systems for diagnostic purposes~\shortcite{adebayo2018sanity}. Following from our previous explanation, the best results are achieved when the highest contour density perfectly overlaps the bright-colored dots. Examples of this kind are Figures~\ref{fig:party-densities-fi},~\ref{fig:party-densities-m5s},~\ref{fig:party-densities-pd},~\ref{fig:party-densities-prc}. Contrarily, many mistakes are made in those cases where regions of bright-colored dots are not contained within any contour line, as in Figures~\ref{fig:party-densities-le} and~\ref{fig:party-densities-e}. In addition to highlighting parties for which we are able to yield good predictions and those for which we are not, Figure~\ref{fig:party-densities} also allows to understand some of the reasons for our mistakes. For example it is evident that for each party, the majority of users is tightly clustered in a restricted area of the ideology space. At the same time however, a minority of users appear to be spread out throughout all the ideology space, possibly also invading a region populated by members of another party. This represents a limitation of our method for learning ideologies or an intrinsic limitation of working with noisy textual data, which inevitably results in wrong predictions at clustering time. For some parties, this drawback is more prominent than for others. For instance, a large portion of \textsf{LE} users completely overlaps the highest-density region of \textsf{FdI}. Such users will be erroneously predicted as supporters of \textsf{FdI} by our technique. Similarly, despite having a high-density cluster that we correctly detected, users of \textsf{+E} also appear to be spread-out across the top-left quadrant of the ideology space, which makes it difficult to cluster them all together. For the future, it will be interesting to evaluate and diagnose novel techniques for learning latent political ideologies and for predicting political leaning, by means of this visualization technique.

\subsubsection{Comparison with Supervised and Semi-Supervised Approaches}
\label{sec:res-supervised}
Results presented in the previous section highlighted the advantages of the proposed \textsf{parties enriched + clustering} technique with respect to all other unsupervised techniques and baselines. However, previous works showed that the additional information exploited by supervised and semi-supervised techniques (e.g., ground-truth labels of the training-set) typically allow to yield better prediction performance compared to unsupervised approaches. Such performance is however hardly generalizable, since it strongly depends on the training set used for learning models. As a consequence, performances reported for supervised and semi-supervised techniques often represent overestimations of the capability to predict political leaning in the wild~\shortcite{cohen2013classifying,yan2019congressional}.

\begin{table*}[t]
    \footnotesize
    \centering
    \begin{adjustbox}{max width=\textwidth}
	\begin{tabular}{llcrrrcrrr}
        \toprule
        \multicolumn{2}{c}{\textbf{method}} && \multicolumn{3}{c}{\textbf{macro}} && \multicolumn{3}{c}{\textbf{micro}} \\
        \cmidrule{1-2}\cmidrule{4-6}\cmidrule{8-10}
        \textit{ideologies} & \textit{predictions} & \textit{type} & \textit{precision} & \textit{recall} & \textit{F1} && \textit{precision} & \textit{recall} & \textit{F1} \\
    	\midrule
    	\textsf{parties enriched} & \textsf{clustering} & \unsupervised     & 0.472 & 0.434 & 0.426 && 0.517 & 0.421 & 0.426 \\
    	\midrule
    	\textsf{supervised enriched} & \textsf{clustering} & \semisupervised       & 0.433 & 0.394 & 0.392 && 0.485 & 0.384 & 0.411 \\
    	\textsf{parties enriched} & \textsf{SVC} & \semisupervised & 0.555 & 0.453 & 0.474 && 0.532 & 0.513 & 0.500 \\
    	\textsf{word2vec} & \textsf{SVC} & \semisupervised & 0.601 & \textbf{0.468} & \textbf{0.485} && \textbf{0.574} & \textbf{0.554} & \textbf{0.536} \\
    	-- & \textsf{majority classifier} & \supervised       & 0.027 & 0.125 & 0.044 && 0.046 & 0.215 & 0.076 \\
    	\textsf{supervised enriched} & \textsf{SVC} & \supervised & \textbf{0.606} & 0.453 & 0.481 && 0.551 & 0.517 & 0.504 \\
    	\bottomrule
    	\multicolumn{9}{l}{\scriptsize \unsupervised~: unsupervised \hspace{0.05\textwidth}\semisupervised~: semi-supervised \hspace{0.05\textwidth}\supervised~: supervised}
	\end{tabular}
	\end{adjustbox}
    \caption{Performance comparison of the \textit{best unsupervised} method against \textit{semi-supervised} and \textit{supervised} ones, for fine-grained (party) prediction of political leaning. The best result for each evaluation metric is shown in bold font.}
	\label{tab:party_supervised_results}
\end{table*}

\begin{table*}[t]
    \footnotesize
    \centering
    \begin{adjustbox}{max width=\textwidth}
	\begin{tabular}{llcrrrcrrr}
        \toprule
        \multicolumn{2}{c}{\textbf{method}} && \multicolumn{3}{c}{\textbf{macro}} && \multicolumn{3}{c}{\textbf{micro}} \\
        \cmidrule{1-2}\cmidrule{4-6}\cmidrule{8-10}
        \textit{ideologies} & \textit{predictions} & \textit{type} & \textit{precision} & \textit{recall} & \textit{F1} && \textit{precision} & \textit{recall} & \textit{F1} \\
    	\midrule
    	\textsf{parties enriched} & \textsf{clustering} & \unsupervised     & 0.751 & 0.752 & 0.750 && 0.776 & 0.772 & 0.772 \\
    	\midrule
    	\textsf{supervised enriched} & \textsf{clustering} & \semisupervised       & 0.748 & 0.745 & 0.745 && 0.787 & 0.785 & 0.785 \\
    	\textsf{parties enriched} & \textsf{SVC} & \semisupervised & 0.828 & 0.769 & 0.789 &&  0.821 & 0.819 & 0.816 \\
    	\textsf{word2vec} & \textsf{SVC} & \semisupervised & \textbf{0.877} & \textbf{0.822} & \textbf{0.841} && \textbf{0.875} & \textbf{0.875} & \textbf{0.871} \\
    	-- & \textsf{majority classifier} & \supervised       & 0.131 & 0.333 & 0.188 && 0.156 & 0.395 & 0.223 \\
    	\textsf{supervised enriched} & \textsf{SVC} & \supervised & 0.822 & 0.761 & 0.780 && 0.822 & 0.823 & 0.816 \\
    	\bottomrule
    	\multicolumn{9}{l}{\scriptsize \unsupervised~: unsupervised \hspace{0.05\textwidth}\semisupervised~: semi-supervised \hspace{0.05\textwidth}\supervised~: supervised}
	\end{tabular}
	\end{adjustbox}
    \caption{Performance comparison of the \textit{best unsupervised} method against \textit{semi-supervised} and \textit{supervised} ones, for coarse-grained (pole) prediction of political leaning. The best result for each evaluation metric is shown in bold font.}
	\label{tab:pole_supervised_results}
\end{table*}

Following this previous line of research, here we are interested in evaluating the performance gap between the best unsupervised technique (\textsf{parties enriched + clustering}) and semi-supervised and supervised ones. Table~\ref{tab:party_supervised_results} shows results of this comparison for the fine-grained prediction task, while Table~\ref{tab:pole_supervised_results} presents results for the coarse-grained task. Results presented in both tables confirm previous findings and show that the best unsupervised technique is outperformed by the best supervised and semi-supervised ones. The best overall results are achieved by the semi-supervised \textsf{word2vec + SVC} technique, with \textit{micro F1} $= 0.536$ on the fine-grained task and \textit{micro F1} $= 0.871$ on the coarse-grained one. Macro results are only slightly worse in both tasks. Thus, the performance gap between the best unsupervised technique and the best overall technique is in the region of $0.11$ \textit{micro F1} on the fine-grained task and $0.10$ \textit{micro F1} on the coarse-grained one. Taking into account the differences previously reported in Tables~\ref{tab:party_unsupervised_results} and~\ref{tab:pole_unsupervised_results} between unsupervised techniques, these last results represent non-negligible yet modest differences in performance. Notably, the \textsf{parties enriched + clustering} unsupervised technique is also capable of beating the \textsf{supervised enriched + clustering} technique in the challenging fine-grained task, in addition to largely beating the simple \textsf{majority classifier} supervised baseline in both tasks.

An interesting result that clearly emerges from Tables~\ref{tab:party_supervised_results} and~\ref{tab:pole_supervised_results} is the superiority of all the approaches based on SVC classifiers for the prediction step. Independently on the methodology used for obtaining political ideologies and on the overall approach to the task (e.g., semi-supervised or supervised), the 3 methods leveraging an SVC consistently obtained the 3 best overall results in both the fine- and coarse-grained tasks. An important contribution to these positive results is given by the data distribution of the different splits of our dataset. In fact, as anticipated in Section~\ref{sec:prelim-dataset}, our data splitting strategy implied that no drift is present between the training and test partitions of our dataset. Under this favorable laboratory condition, supervised classifiers are able to maximize their learning phase on data instances in the training-set, and to effectively carry over what they learnt to the test-set. However, it is known that real-world conditions are characterized by issues such as concept drift that limit the generalizability of supervised approaches~\shortcite{lu2018learning}. In presence of concept drift, or of any other factor that shifts the test distribution away from the one used in training, supervised approaches end up being unreliable. Instead, unsupervised approaches, such as the one proposed in our work, are able to better adapt to possible drifts. For instance, with reference to the ideology space shown in Figure~\ref{fig:party-densities}, while a supervised classifier learns fixed decision boundaries for the different parties based on the data distribution of the training-set, our unsupervised clustering approach is capable of highlighting regions of the ideology space featuring high density, independently on their position.

\subsubsection{Validation: Concept Drift}
The results presented so far are computed on the test-set split of our dataset, obtained via a stratified random sampling of the users as explained in Section~\ref{sec:prelim-dataset}. However, as anticipated in the previous section, a more rigorous evaluation can be conducted by assessing the performance of our technique on a time-dependent test-set, by assigning users to either the training-, validation-, or test-set according to the time when they tweeted. The advantage of this evaluation strategy is that, in general, time-wise splits are more representative of the conditions in which machine learning models are used, as they allow to test a model’s ability to withstand issues that emerge through time, such as concept-drift~\shortcite{lu2018learning}. In turn, a model capable of withstanding such issues would open up the possibility to carry out longitudinal analyses and even to nowcast political leanings~\shortcite{lampos2012nowcasting,avvenuti2017nowcasting,tsakalidis2018nowcasting}. For these reasons, we performed an additional experiment by evaluating our method in this, more stringent, condition.

Specifically, we first obtained a new time-wise test-set that contains 5,524 users that only tweeted after August 15, 2019. All other users from our dataset are assigned to either the training- or validation-set, which contain users that tweeted before the threshold date. Next, we used our best method (i.e., \textsf{parties enriched + clustering}) to repeat both the party (fine-grained) and the pole (coarse-grained) prediction tasks. Finally, we compared the results obtained by our method on the time-dependent test-set with those obtained on the original (random) one. Regarding party predictions, our method obtains \textit{macro F1} $= 0.388$ and \textit{micro F1} $= 0.389$ on the time-dependent test-set, whereas it obtained both \textit{macro} and \textit{micro F1} $= 0.426$ on the random one. For pole predictions, our method obtains \textit{macro F1} $= 0.721$ and \textit{micro F1} $= 0.771$ on the time-dependent test-set, whereas it obtained \textit{macro F1} $= 0.750$ and \textit{micro F1} $= 0.772$ on the random one. Summarizing, the more stringent evaluation resulted in a maximum of $9\%$ performance decrease on the party prediction task, and in a maximum of $4\%$ performance decrease on the pole prediction one. The overall positive results of our unsupervised method are confirmed.

\subsubsection{Validation: Members of the European Parliament}
The ground-truth labels for users of our dataset are implicitly derived from user likes to party tweets, as detailed in Section~\ref{sec:prelim-ground-truth}. On the one hand this labeling strategy removed the need for a manual labeling of our dataset and avoided possible human errors and biases entering our ground-truth~\shortcite{pandey2019modeling}. On the other hand however, an automatic labeling strategy does not necessarily exclude the risk of inconsistencies or wrong labels. In order to further validate the correctness of our predictions, we also applied our technique to a small set of users whose preferred party and pole are publicly known. Specifically, we focused on the Italian members of the European parliament (MEPs) as of 2019. These represent active politicians whose party and pole affiliations can be used as reliable ground-truth labels in our prediction tasks. Notably, the majority of MEPs also has an official Twitter account linked to its public page on the European parliament website\footnote{\url{https://www.europarl.europa.eu/meps/en/home}}.

Similarly to the users in our Twitter dataset, we thus collected 200 tweets for each of the 34 Italian MEPs with an official Twitter account. Then, we applied our best method (i.e., \textsf{parties enriched + clustering}) to predict the party leaning of the MEPs. Finally, we evaluated our method by comparing its predictions with the party affiliations of the MEPs. We repeated these steps also for the method by~\shortciteA{darwish2020unsupervised}, thus enabling a comparison of our results with those of this state-of-the-art technique. On this party prediction task, our method achieved \textit{macro F1} $= 0.651$ and \textit{micro F1} $= 0.823$. Instead, the method by~\shortciteA{darwish2020unsupervised} achieved \textit{macro F1} $= 0.377$ and \textit{micro F1} $= 0.662$. Overall, our results are very positive, confirming the good performance of our technique and its superiority with respect to the best existing unsupervised competitor. We also note that the results reported in this section are better than those reported in Tables~\ref{tab:party_unsupervised_results} and~\ref{tab:party_supervised_results}. This is expected by considering that MEPs are very politically-active users with clear political inclinations. As such predicting their political leaning represents a simpler task with respect to predicting the political leaning of generic Twitter users.

\begin{figure*}[t]
    \centering
    \includegraphics[width=0.65\textwidth]{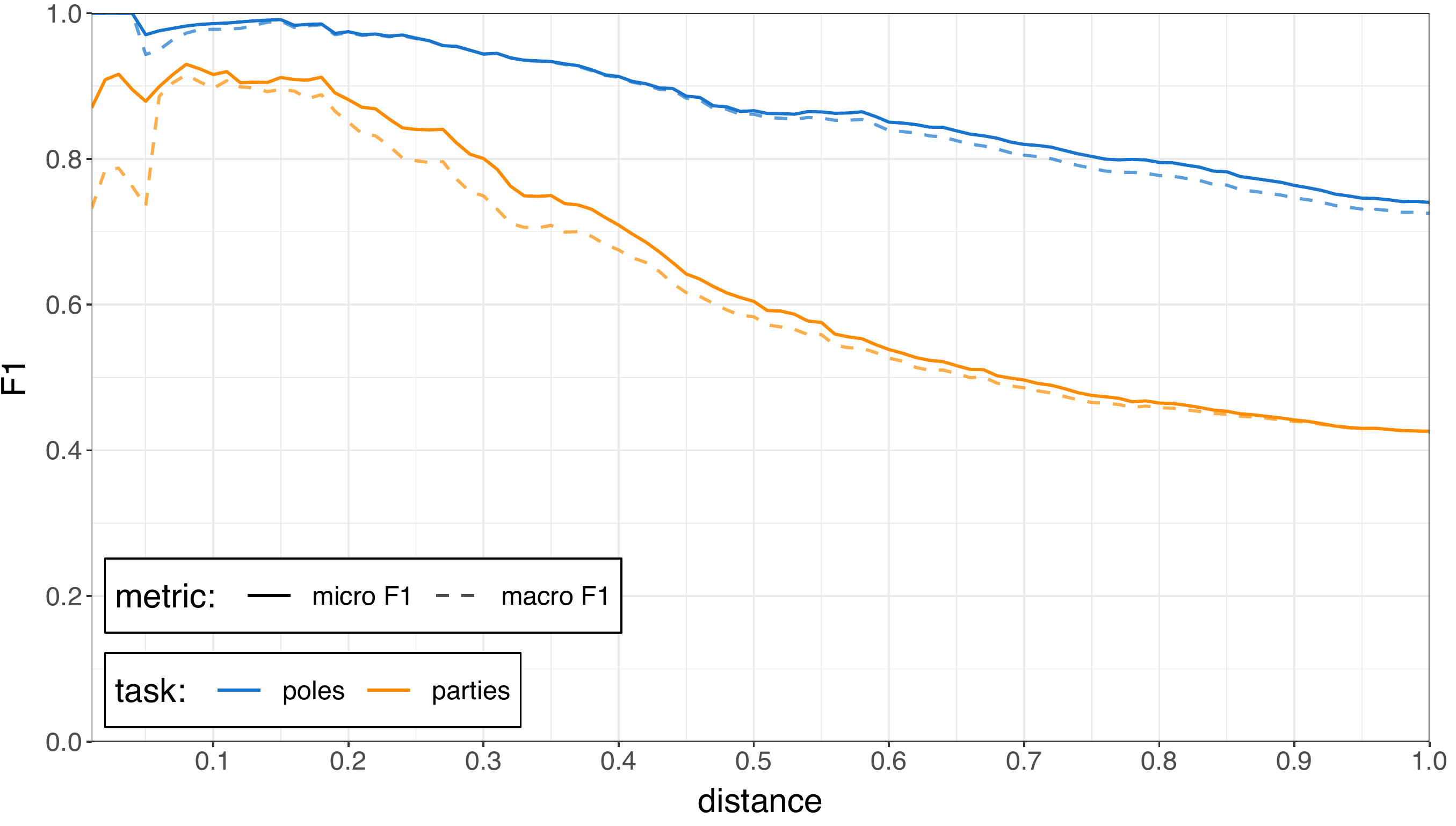}
    \caption{Performance evaluation of our \textsf{parties enriched + clustering} unsupervised technique as a function of user distance from the pivots. For users that lay near to one of the pivots, we are able to provide fine- (party) and coarse-grained (pole) predictions with exceptional accuracy. Instead, most of our mistakes occur for users that are positioned far away from all pivots.}
    \label{fig:results-distance}
\end{figure*}

\subsubsection{Sensitivity Analysis: Distance}
\label{sec:res-distance}
In this section and in the subsequent ones, we report results of a sensitivity analysis that we carried out on the best unsupervised technique that we proposed: \textsf{parties enriched + clustering}.

The rationale for the analysis discussed in this section stems from results of Tables~\ref{tab:party_unsupervised_results} and~\ref{tab:pole_unsupervised_results}. In particular, the evaluation of the unsupervised approaches highlighted the promising performance of the \textsf{parties enriched + distance} baseline. Given an ideology space, this technique assigns a label to each user based on its distance to the pivots. In other words, the simple distance-based prediction strategy employed in this baseline was capable of yielding positive results. This suggests that the distance from the pivots in the ideology space is a relevant parameter that has an impact on our predictions. With the goal of evaluating this facet, in Figure~\ref{fig:results-distance} we show results of an analysis where we evaluated the performance of our \textsf{parties enriched + clustering} technique, as a function of user distance from the pivots. Results shown in figure confirm our previous intuition. When only evaluating predictions for users that lay near to one of the pivots, our results are extremely accurate. For instance, when considering only users whose min-max normalized distance $\leq 0.2$, our technique obtains \textit{micro F1} $> 0.90$ and $> 0.98$ on the fine- and coarse-grained task, respectively. Exceptional results indeed, considering the difficulty of the tasks at hand. As we include in our evaluation also users who lay further away from any pivot, our results worsen. At the end of our evaluation, when we consider all users independently on their distance, we end up with the same results already reported in Tables~\ref{tab:party_unsupervised_results} and~\ref{tab:pole_unsupervised_results}.

The decreasing trend shown in Figure~\ref{fig:results-distance} demonstrates that the accuracy of our predictions strongly depends on a user's distance to the pivots. In turn, this facet can be exploited to complement our predictions with a confidence score that states how likely a given prediction is to be correct. For users that lay near to one of our pivots in the ideology space, we are able to provide predictions with large confidence scores (e.g., $>90\%$). Conversely, for users that lay far away from all pivots, we are still able to provide a prediction, but with a much lower confidence.

These results also suggest one possible strategy for quickly improving the performance of our technique -- that is, increasing the number of pivots. This simple operation would reduce the average distance of users from the pivots, thus allowing to obtain overall better performances. However, having a large number of pivots requires additional manual effort and it would also move the approach towards a semi-supervised one, with all the implications previously discussed. We remark that for all the experiments in this work, we used the minimum possible number of pivots: one for each considered political party.

\begin{figure*}[t]
    \centering
    \includegraphics[width=0.65\textwidth]{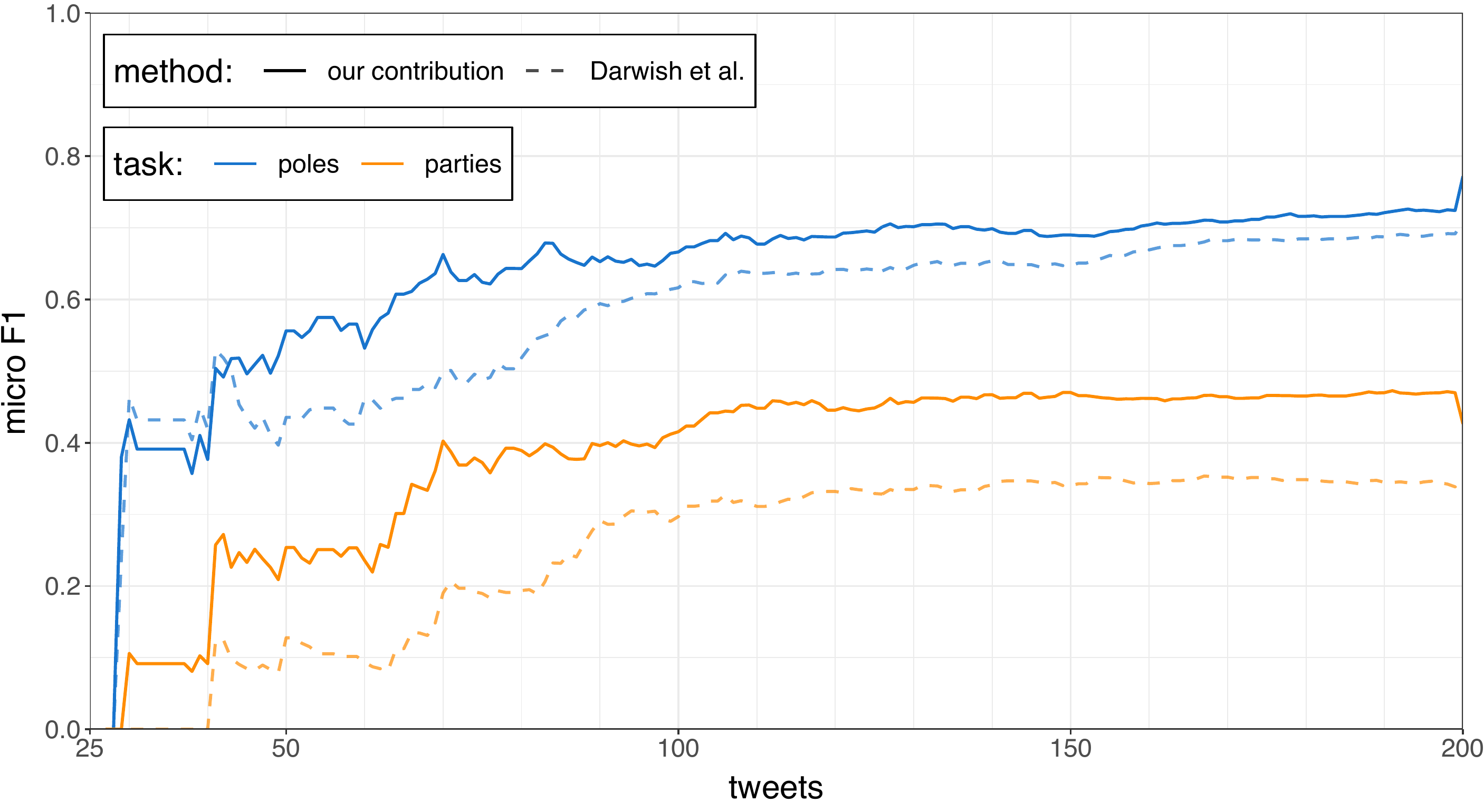}
    \caption{Performance evaluation of~\shortciteA{darwish2020unsupervised} and of our \textsf{parties enriched + clustering} unsupervised technique, as a function of the number of user tweets. Our technique consistently beats the competitor, especially in the challenging party prediction (i.e., fine-grained) task. Both techniques show overall stable performance for users having $\ge 100$ tweets. For users having $< 100$ tweets the performance of both techniques starts decreasing. For $< 40$ tweets performances rapidly plummet.}
    \label{fig:results-tweets}
\end{figure*}

\subsubsection{Sensitivity Analysis: Tweets}
\label{sec:res-tweets}
Our proposed technique is based on the analysis of the textual content of the tweets shared by social media users. One important aspect to consider when evaluating our technique is thus its sensitivity to the number of available tweets per user. Intuitively, users for which only a small number of tweets are available represent more challenging predictions than users who shared many tweets. With the goal of evaluating this aspect, Figure~\ref{fig:results-tweets} shows the performance of our technique and of~\shortciteA{darwish2020unsupervised}, as a function of the number of user tweets, for both tasks. We recall that we collected maximum 200 tweets per user and that we discarded users for which we could collect $< 25$ tweets.

Results in Figure~\ref{fig:results-tweets} show that our technique consistently beats the competitor in both tasks and for all users, independently on the number of their tweets. The only exception is represented by users having $< 40$ tweets, on the fine-grained party prediction task, for which~\shortciteauthor{darwish2020unsupervised} obtain slightly better results than us. Apart from this, our method always achieves superior results, especially in the challenging party prediction task. Interestingly, both techniques show overall stable performance for users having $\ge 100$ tweets. Instead, as expected, for users having $< 100$ tweets the performance of both techniques starts decreasing. The decreasing trend is particularly steep for users having $< 40$ tweets, for which the performances of both techniques rapidly plummet.

In addition to serving as further evaluation of our technique, these results also provide guidance for applying tweet-based predictors of political leaning in-the-wild. In fact, our experiments suggest that near-optimal results can be expected for users with $\ge 100$ tweets, and reduced performance otherwise. In particular, predictions obtained for users having $< 40$ or $50$ tweets should undergo additional scrutiny and validation, since misclassifications are frequent under these operating conditions.

\begin{figure*}[t]
    \centering
    \includegraphics[width=0.65\textwidth]{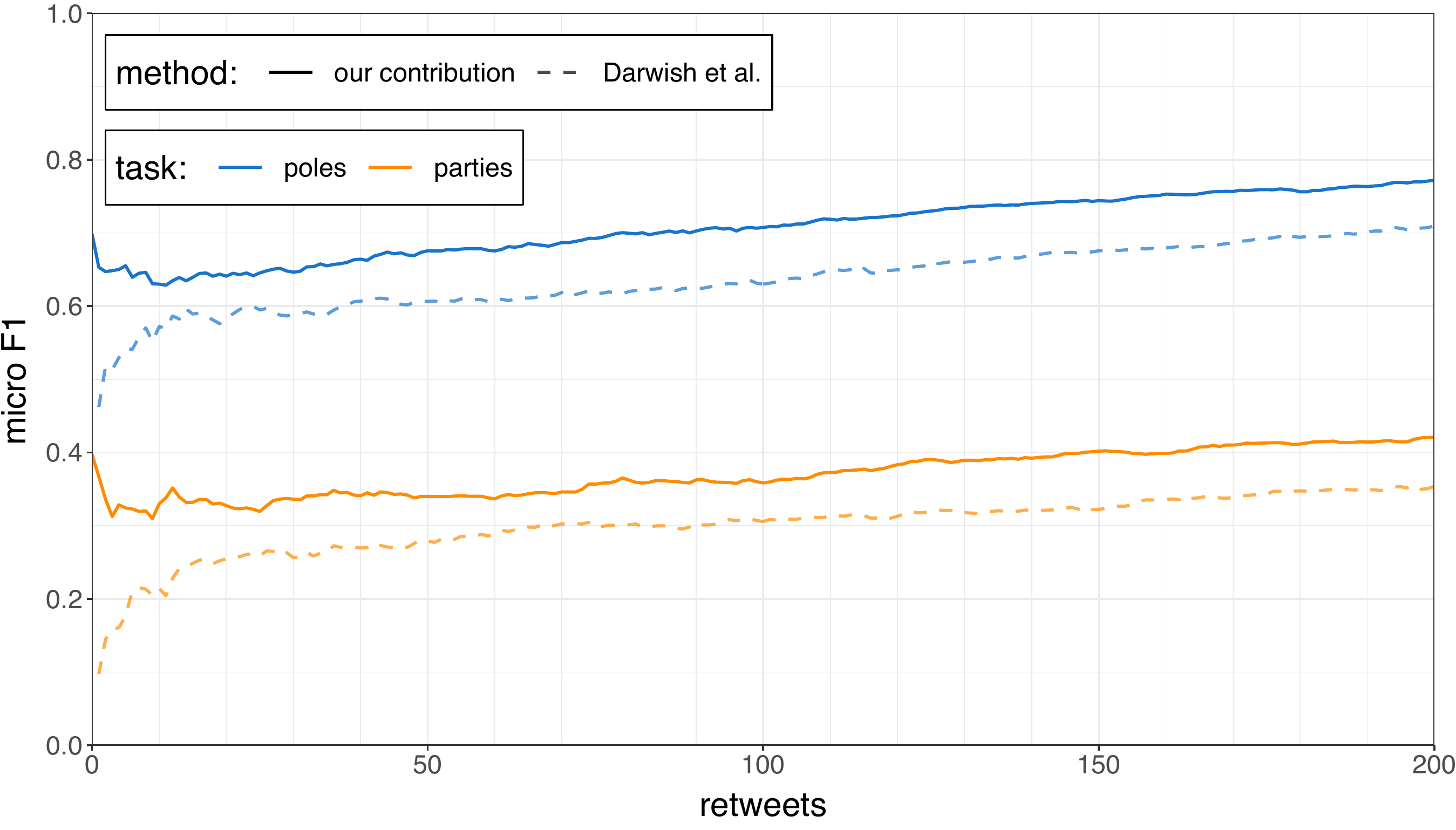}
    \caption{Performance evaluation of~\shortciteA{darwish2020unsupervised} and of our \textsf{parties enriched + clustering} unsupervised technique, as a function of the number of user retweets. Both techniques show degraded performance when classifying users with few retweets, showing the importance of this signal for the task. However, our technique consistently outperforms the competitor and does not suffer from a steep performance drop for users with $\leq 15$ retweets.}
    \label{fig:results-retweets}
\end{figure*}

\subsubsection{Sensitivity Analysis: Retweets}
\label{sec:res-retweets}
In many recent works, retweets have been used as a strong signal in several prediction tasks on social media, including the estimation of stance and political leaning~\shortcite{aldayel2019your}, degree of automation~\shortcite{mazza2019rtbust}, extent of coordination among users~\shortcite{nizzoli2020coordinated} and percentage of fake messages in an online discussion~\shortcite{tardelli2021detecting}, to name but a few notable examples. In particular, one of the techniques that we evaluated in this work solely depends on retweets for estimating political leaning~\shortcite{darwish2020unsupervised}. In addition, also our proposed technique uses that information, although not as explicitly as by~\shortciteA{darwish2020unsupervised}. In fact, as explained in Section~\ref{sec:ideology}, in our work retweets only partly contribute to document embeddings, which in turn contribute to our latent user representations.

Similarly to the previous experiment, here we were interested in evaluating the impact that retweets have on the predictions of political leaning generated by our technique and by that of~\shortciteA{darwish2020unsupervised}. To carry out this experiment, we repeatedly evaluated both techniques on subsets of test-set users featuring different numbers of retweets, starting from users with no retweets at all, and concluding our experiment with users with 200 retweets (the maximum number of tweets that we collected per each user). Results are shown in Figure~\ref{fig:results-retweets}. As expected, both techniques achieve worse results for users with few retweets, confirming the informativeness of this feature. Our proposed technique consistently outperforms the one from~\shortciteA{darwish2020unsupervised} and the gap between the 2 shows only minor fluctuations along the \textit{x} axis. However, a marked difference is shown for users that feature an extremely low number of retweets. Indeed for users with $\leq 15$ retweets, the performance of~\shortciteA{darwish2020unsupervised} plummets in both prediction tasks. On the contrary, our technique exhibits a different behavior, as it does not appear to be impacted so negatively by an extremely low number of retweets. The difference between the behavior of the 2 techniques is explained by considering that retweets are the only information exploited by~\shortciteA{darwish2020unsupervised}, while they are an important -- yet minor -- part of all the information that our technique leverages.

\subsubsection{Limitations and Open Challenges}
\label{sec:limitations}
In this section we carry out a detailed analysis of the main types of errors made by our proposed method. To reach this goal, we manually selected a set of users that were projected by our technique to a region of the latent ideology space that is not associated with their ground-truth party label. For these users that were projected far from their party -- and thus, that were subsequently wrongly labeled by the clustering step -- we manually analyzed their Twitter timelines, so as to identify the root causes for our misclassifications. This analysis contributes to highlighting current limitations in content-based unsupervised approaches to the prediction of social media political leaning, also highlighting open challenges and valuable directions for future research.

\begin{table*}[t]
    \footnotesize
    \centering
    \begin{adjustbox}{max width=\textwidth}
	\begin{tabular}{p{6cm}p{6cm}cc}
        \toprule
        \textbf{original tweet} & \textbf{translated tweet} & \textbf{party} & \textbf{score} \\
        \midrule
        \multicolumn{4}{l}{\textit{Type 1: tweets in favour of a party/politician, that receive low scores for that party:}} \\ [0.7em]
        @matteosalvinimi Matteo Salvini gli italiani sceglieranno il miglior Matteo (Salvini) & @matteosalvinimi Matteo Salvini Italians will choose the best Matteo (Salvini) & \textsf{LE} & 0.126\\
        \hdashline
        @LegaSalvini Mitico Matteo Salvini sei il nostro capitano & @LegaSalvini Mythical Matteo Salvini you are our captain & \textsf{LE} & 0.066\\
        \midrule
        \multicolumn{4}{l}{\textit{Type 2: tweets against a party/politician, that receive high scores for that party:}} \\ [0.7em]
        Deve andare in pensione. Berlusconi ormai è fulminato. & He has to retire. Berlusconi is stoned. & \textsf{FI} & 0.454\\
        \hdashline
        @DSantanche @FratellidItaIia tornare? devi cominciare a crescere Santanchè, hai novant'anni e "ragioni" come una lattante & @DSantanche @FratellidItaIia coming back? you have to grow Santanchè, you are ninety years old and you still "think" like a baby & \textsf{FdI} & 0.470\\
        \midrule
        \multicolumn{4}{l}{\textit{Type 3: tweets with limited/no political information, that receive high scores for a party:}} \\ [0.7em]
        RT @oss\_romano: \#27agosto \#rassegnastampa Un mondo di fraternità e pace è possibile. Il \#Papa incoraggia le iniziative per dare attuazione ... & RT @oss\_romano: \# 27agosto \#rassegnastampa A world of fraternity and peace is possible. The \#Pope encourages initiatives to implement ... & \textsf{PRC} & 0.658\\
        \hdashline
        RT @visit\_lazio: Tra le 100 esperienze al mondo da vivere, il settimanale @TIME include il @Castello\_Severa nella lista world’s greatest ... & RT @visit\_lazio: Among the 100 experiences in the world to live, the @TIME magazine includes @Castello\_Severa in the world’s greatest ... & \textsf{M5S} & 0.480 \\
        \midrule
        \multicolumn{4}{l}{\textit{Type 4: tweets with local, subjective, or very specific information, that receive high scores for a party:}} \\ [0.7em]
        RT @c\_appendino: Asfalto nuovo per via Cigna. Una buona notizia per i tanti cittadini che transitano su questa importante arteria & RT @c\_appendino: New asphalt on via Cigna. Good news for the many citizens who move through this important thoroughfare & \textsf{M5S} & 0.773\\
        \hdashline
        RT @virginiaraggi: Partiti i lavori di restauro della Fontana delle Rane nel quartiere Coppedè. L'intervento è il primo di questa portata & RT @virginiaraggi: Restoration work on the Fontana delle Rane in the Coppedè district has begun. The intervention is the first of this magnitude & \textsf{M5S} & 0.602\\
    	\bottomrule
	\end{tabular}
	\end{adjustbox}
    \caption{Examples of problematic tweets that were incorrectly assessed by our tweet party classifier. For each of the 4 main types of problematic tweets, we report some examples specifying the reference political party to which the error is referred and the corresponding score computed by the tweet party classifier. Scores are in the $[0, 1]$ range.}
	\label{tab:wrong_classified_tweets}
\end{table*}

We first evaluated \textit{major} misclassifications -- namely, cases where users were projected to a region of the ideology space related to parties of the opposite pole with respect to the ground-truth party of the users. For example, users favoring a left-leaning party (e.g., \textsf{PD}) that were erroneously projected to a region of the ideology space associated with extreme-right parties (e.g., \textsf{LE}, \textsf{FdI}, \textsf{CPI}). These cases yield errors both in the fine-grained party prediction task, as well as in the coarse-grained pole prediction task. Overall, the total number of these major misclassifications is small, but it is nonetheless interesting to assess the causes for these errors. The analysis of these major misclassifications revealed that some of our ground-truth labels contrast with the information contained in the tweets from the user timeline. Ground-truth labels were automatically obtained from user likes to party tweets. Instead, our classifications are derived from user tweets. Thus, the majority of cases of major misclassifications are related to users that liked many tweets by a given party, but that support a different party in their own tweets. This is a peculiar and interesting behavior that, to the best of our knowledge, is undocumented. The existence of a subset of users exhibiting this behavior mandates to carefully consider the source of ground-truth labels in future works, since considering user likes or following relationships might convey different and contrasting information with respect to that obtainable from user tweets. In the remaining cases of major misclassifications by our system, we were not able to correctly detect the political leaning of the user mainly due to: (i) wrong understanding of tweet semantics (more on this in the following); or (ii) an objective difficulty in understanding the political orientation of the user, due to ambiguous and contrasting political content. This latter case is not a limitation of our technique, since also human evaluators would struggle to reliably provide predictions for certain users, but rather an inherent challenge in the classification of users that express few or ambiguous political positions. Such challenge has already been noted in earlier works on this same task~\shortcite{cohen2013classifying}, as well as on other social media-related tasks~\shortcite{cresci2018real}.

We also assessed causes of \textit{minor} misclassifications -- namely, cases where a user is labeled with a wrong party in the fine-grained party classification task, but it is correctly labeled in the coarse-grained pole prediction task. In such cases, several misclassifications are caused by a wrong interpretation of tweet semantics or by the weight (i.e., the importance) that our system assigned to certain political tweets. 
In fact, the political orientation of a user is not a binary concept, but it is rather a nuanced concept often involving ideas and opinions that align with the political line of more than one party. In particular, it is common for a user to support opinions from multiple politically-close parties. To this regard our system, as a human evaluator would do, weighs the available information as best as possible, but with an inevitable degree of uncertainty. Finally, we also analyzed errors for users projected to the central (i.e., most uncertain) area of the ideology space of Figure~\ref{fig:party-projections-ours}, finding that the projection errors are mainly due to one of the following reasons: (i) users with insufficient political content in their timeline; (ii) automated accounts that produce extremely varied content (i.e., news bots); and (iii) users that repeatedly attack certain political parties and leaders, but that do not explicitly support any party\footnote{For these users, we know who they \textit{do not} support, but we do not know who they \textit{do} support.}. Challenges related to the analysis of the first category of users are well-known in literature. For instance,~\shortciteA{cohen2013classifying} refer to them as \textit{politically inactive} users. Instead, challenges related to the analysis of bots and political antagonists, are rather undocumented, despite the widespread presence of both these types of accounts in our online ecosystems~\shortcite{lokot2016news,nizzoli2020coordinated}.

Given that our projections are based on the scores assigned to user tweets by the tweet party classifier, wrong predictions by our technique are typically due to initial errors by the tweet party classifier. We now turn our attention to these errors, so as to provide practical examples of problematic tweets. Our analysis highlighted 4 main categories of problematic tweets, summarized in Table~\ref{tab:wrong_classified_tweets}. A first set of errors is due to problematic tweets of type 1 (tweets in favour of a party/politician, that receive low scores for that party). Here, our classifier was unable to provide high scores for the correct party because of the limited number of these tweets used to train it. Increasing our dataset, or anyway feeding more tweets of this type to the classifier, would likely remove this type of error. Errors due to the second type of problematic tweets (tweets against a party/politician, that receive high scores for that party) are more challenging. First of all, those tweets do not express support for any party nor candidate. Thus, there is an intrinsic difficulty in assigning a high score for a party. Moreover, they negatively -- yet explicitly -- mention a party, which tricked our classifier into giving a high score for that party. This second issue implies that our deep learning tweet party classifier was unable to correctly ``understand'' the meaning of those tweets. This problem can be mitigated by implementing the classifier with more complex and powerful deep learning architectures, such as those based on modern pretrained language models (e.g., BERT, T5). These state-of-the-art natural language understanding systems are capable of grasping subtle nuances in the language used for or against a given political party. As such, they would contribute to reducing this type of errors. The third type of problematic tweets (tweets with limited/no political information, that receive high scores for a party) are due to the challenges of classifying items that do not convey any useful information for the machine learning task at hand. In this situation, classifiers usually yield unreliable predictions. One possible way of solving this issue is by carrying out an additional filtering step in our analysis pipeline. For instance, we could train a separate binary classifier to distinguish between politically-related and unrelated tweets. Then, only politically-related tweets would be given to the party tweet classifier for computing a party score. The last type of problematic tweets (tweets with local, subjective, or very specific information, that receive high scores for a party) represents another big challenge. In the examples from the bottom rows of Table~\ref{tab:wrong_classified_tweets}, a user is expressing positive opinions about the local administration. Notably, the high scores given by our classifier do not necessarily represent errors, in a strict sense. In fact, appreciation for the work of a local administration surely conveys a certain extent of political information. However, for these users our classifier should have given more weight to other, more explicit, political tweets with respect to those supporting the local administration.

\section{Conclusions}
\label{sec:conc}
We proposed a novel unsupervised technique for estimating the political leaning of social media users. Our solution leverages a deep neural network in a representation learning task, for analyzing user tweets and for learning latent political ideologies. Then, users are projected in a low-dimensional ideology space and are subsequently clustered. The political leaning of a user is automatically derived from the cluster to which the user is assigned. We evaluated our technique on two prediction tasks and we compared it to baselines and other state-of-the-art approaches. The fine-grained task aims to infer the preferred political party of each user, out of 8 possible parties. The -- easier -- coarse-grained task aims to infer the high-level political leaning of each user, in a 3-class classification task. Among all unsupervised techniques that we evaluated, our proposed one achieved the best results in both tasks, with \textit{micro F1} $= 0.426$ and $0.772$, respectively for the fine- and coarse-grained task. It also achieved comparable results to some of the semi-supervised and supervised techniques. However, the best unsupervised technique is outperformed by the best semi-supervised and supervised ones, given the additional information that the latter exploit. Moving forward, we demonstrated that we can exploit the topology of our learned ideology space to assign a confidence score to our predictions, thus allowing to retain only those predictions for which the confidence meets a desired threshold. Finally, we analyzed the relationship between our predictions and the number of tweets and retweets performed by users, showing that our technique is able to provide accurate predictions also for users who tweet or retweet sporadically, contrarily to other state-of-the-art methods.

Our results advance the state-of-the-art for unsupervised prediction of political leaning -- an increasingly popular task. For the future we aim to provide additional contributions by devising better techniques for learning latent political ideologies, a step where there is still large room for improvement. To this regard, another interesting direction of research involves providing interpretations to the dimensions of the latent ideology space. As shown in our results, the different parties seem to position themselves in different regions of the space. Hence, being able to interpret the main dimensions of the ideology space could provide additional and valuable information. For the future we also aim at learning aspect-based stances on a number of politically-relevant issues (e.g., immigration, economy, rights, and more). Finally, we plan to leverage our technique for carrying out a longitudinal analysis aimed at investigating fluctuations in leaning due to important real-world events, such as during electoral campaigns. This latter experimental setup would also be valuable toward assessing the robustness of our system, as well as of others tackling the same task, to known issues such as concept drift and other temporal variations.

\acks{Correspondence should be addressed to Dr. Stefano Cresci (\textit{stefano.cresci@iit.cnr.it}).This research is supported in part by the EU H2020 Program under the scheme ``INFRAIA-01-2018-2019: Research and Innovation Action'' grant agreement \#871042 \textit{SoBigData++: European Integrated Infrastructure for Social Mining and Big Data Analytics}.}

\vskip 0.2in
\bibliography{references}
\bibliographystyle{theapa}

\end{document}